\documentclass[aps,prf,showpacs,superscriptaddress,groupedaddress]{revtex4}  
\usepackage{graphicx,rotating,color}  
\usepackage{bm}        
\usepackage{amssymb}   
\usepackage{lineno,hyperref}
\usepackage{epstopdf}
\usepackage{mathptmx}
\usepackage{amsmath}
\usepackage{caption}
\usepackage{subcaption}
\usepackage{multirow}
\usepackage{hyperref}
 \hypersetup{
     colorlinks=true,
     linkcolor=blue,
     filecolor=blue,
     citecolor = blue,
     urlcolor=cyan,
     }
\usepackage{comment}
\usepackage{lipsum}
\usepackage{soul}

\begin{document}

\title{Surfactant-laden bubble bursting: dynamics of capillary waves and Worthington jet \\ at large Bond number}

\author{P. Pico}
\affiliation{Department of Chemical Engineering, Imperial College London, South Kensington Campus, London SW7 2AZ, United Kingdom}
\author{L.~Kahouadji}
\email{l.kahouadji@imperial.ac.uk}
\affiliation{Department of Chemical Engineering, Imperial College London, South Kensington Campus, London SW7 2AZ, United Kingdom}
\author{S.~Shin}
\affiliation{Department of Mechanical and System Design Engineering, Hongik University, 04066 Seoul, Korea}
\author{J.~Chergui} 
\affiliation{Universit\'e Paris Saclay, Centre National de la Recherche Scientifique (CNRS), Laboratoire Interdisciplinaire des Sciences du Num\'erique (LISN), 91400 Orsay, France}
\author{D.~Juric}
\affiliation{Universit\'e Paris Saclay, Centre National de la Recherche Scientifique (CNRS), Laboratoire Interdisciplinaire des Sciences du Num\'erique (LISN), 91400 Orsay, France}
\affiliation{Department of Applied Mathematics and Theoretical Physics, University of Cambridge,  Cambridge CB3 0WA, United Kingdom}
\author{O.~K.~Matar}
\affiliation{Department of Chemical Engineering, Imperial College London, South Kensington Campus, London SW7 2AZ, United Kingdom}

\date{\today}

\begin{abstract}
We present a numerical study of the main sub-stages preceding aerosol formation via bursting bubbles: capillary wave propagation along the bubble, convergence at the bubble's apex, the ascent of a Worthington jet and its break-up to release liquid drops. We focus on two crucial yet overlooked aspects of the system: the presence of surface-active agents and dynamics driven by non-negligible gravitational effects, quantified by the Bond number. Our results propose, for the first time, a mechanism explaining capillary wave retardation in the presence of surfactants, involving the transition from bi- to uni-directional Marangoni stresses, which pull the interface upwards, countering the motion of the waves. We also quantitatively elucidate the variable nature of the waves' velocity with various surfactant parameters, including surfactant solubility and elasticity, a departure from the constant behaviour well-documented in clean interfaces.
\end{abstract}

\maketitle


\section{Introduction}\label{sec:introduction}

Aerosol droplets, arising from the bursting of bubbles at a liquid-gas interface, hold a crucial position at the confluence of diverse natural phenomena, industrial applications, and daily activities. Consider, for instance, the effervescence in sparkling drinks like champagne, where the bubbles that nucleate at the glass walls and eventually burst at the surface release the aromas experienced by the human senses \cite{Ghabache2016SizeBursting}. In atmospheric physics, marine aerosol plays a pivotal role in governing the overall mass and heat exchange process between the oceans and the atmosphere. This stems from the liberated liquid drops, formed from an estimated $10^{18} - 10^{20}$ bursting bubbles per second across the oceans \cite{Seon2017EffervescenceEvaporation} and capable of transporting an array of often surface-active substances, including salts, dissolved gases, biological material (e.g., harmful pathogens), microplastics, and toxic chemicals. Spanning a broad size range, from a few nanometers to several millimetres \cite{Veron2015OceanSpray}, marine aerosol actively participates in atmospheric chemical reactions, acts as nuclei for cloud and fog condensation, and can influence the cyclone intensity \cite{Veron2015OceanSpray,Deike2018DynamicsBubbles}. 

\begin{figure}
\includegraphics[width=1.0\linewidth]{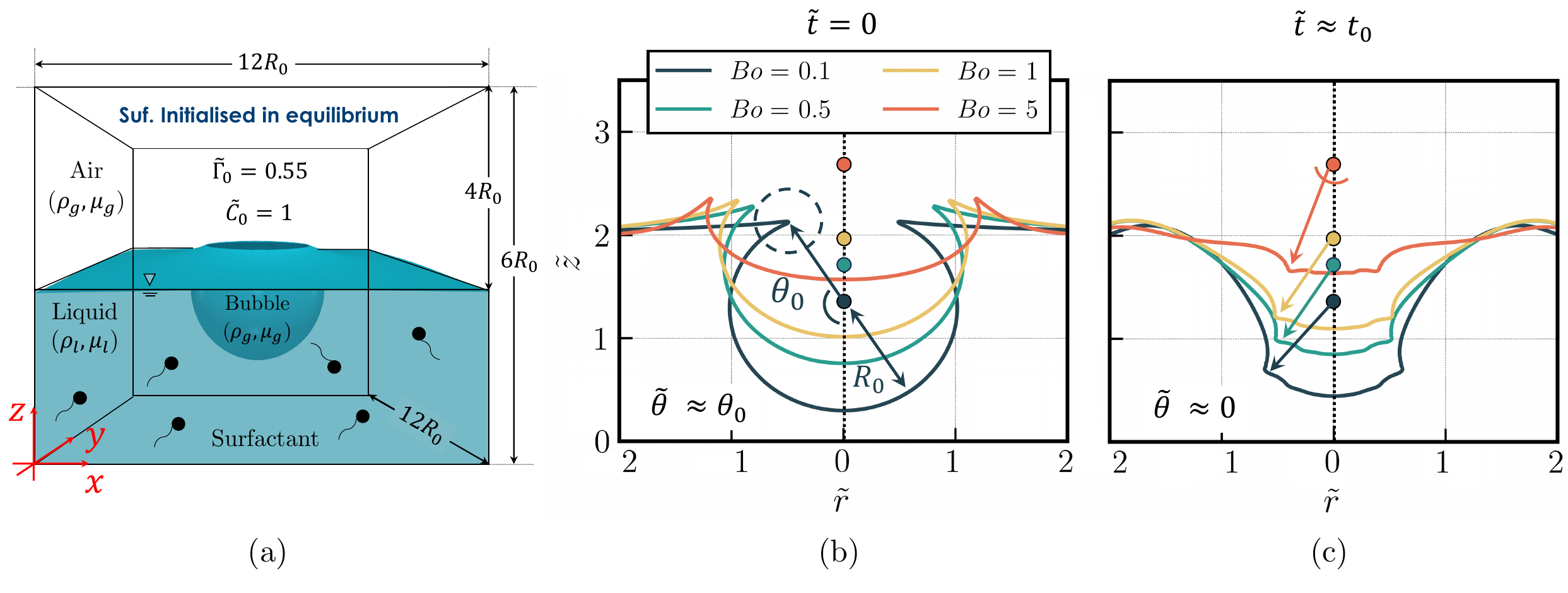}
  \caption{3-D computational domain in a Cartesian coordinate system $\textbf{x} = (x,y,z)$ and initial bubble shapes. A radial coordinate is defined as $r = \sqrt{(x - x_{0})^{2} + (y - y_{0})^{2}}$, where $x_{0}$ and $y_{0}$ correspond to the domain's centre point. (a) Schematic representation of the simulated domain, showcasing an equilibrated initial bubble immersed in a liquid pool with initial surfactant concentration $\tilde{C}_{0}$ and interfacial concentration $\tilde{\Gamma}_{0}$ (not to scale). (b) 2-D projection in a $x-z$ plane of initial ($\tilde{t} = 0$) bubble shape for different $Bo$. The points located at $\tilde{r} = 0$ mark the centre of each bubble, from which the angle of the dominant capillary wave is measured \cite{Gordillo2019CapillaryJets}. (c) Representation of an instant approaching cavity collapse ($\tilde{t} \approx t_{0}$) at the bubble apex, where $\tilde{\theta} \approx 0$ }
  \label{fig:sim_conf}
\end{figure}

The aforementioned significance of aerosol droplets generated via bursting of bubbles has made the system a notable problem within the realm of fluid mechanics, witnessing substantial advancements since the latter half of the previous century. In the absence of contaminants, the dynamics of an isolated bursting bubble of radius $R_{0}$ are governed by the interplay amongst buoyancy, surface tension, and viscosity, quantified by the Laplace, $La = \rho_{l}\sigma_{s}R_{0}/\mu^{2}_{l}$ (which can also be defined as Ohnesorge, $Oh = La^{-1/2}$) and the Bond, $Bo = \rho_{l} g R^{2}_{0} / \sigma_{s}$, numbers. In these equations, $\rho$, $\mu$, $\sigma_{s}$, and $g$ correspond to density, viscosity, surface tension, and gravity, respectively, and the subscript $l$ represents the continuous liquid phase. A typical initiation of an isolated bursting event involves a spherical bubble rising in a liquid medium, reaching the surface, and creating a thin liquid film (characterised by a length scale $\delta / R_{0} \sim O(10^{-6}) - O(10^{-3})$) that separates the bubble from the atmosphere. In situations where $R_{0} < l_{c}$, $l_{c}$ being the capillary length scale \cite{Villermaux2022BubblesMysteries,Lhuissier2012BurstingAerosols}, this thin film undergoes drainage at an exponential or algebraic rate \cite{Yu2023BubbleReview} until its ultimate collapse, triggering the development of capillary waves that propagate on two fronts: across the bubble's domain, and away from the cavity (see figure \ref{fig:sim_conf}(a)). The bubble's capillary waves converge at the cavity's origin and induce a far-field flow which accelerates the liquid phase towards the axis of symmetry \cite{Gordillo2023TheoryJets} and causes a reversal of the bubble-liquid interface, bringing about the ejection of a high-velocity (up to 50 times the capillary velocity, $u_{c} = (\sigma_{s} / (\rho_{l}R_{0}))^{1/2}$) Worthington jet. Depending on the conditions, this jet may break up via a Rayleigh–Plateau instability to release one, or multiple, liquid drops into the atmosphere.

The fundamental dynamics governing the formation of jet drops from bursting bubbles are well-established in scenarios wherein the system is devoid of any contaminant material and buoyancy plays a negligible role ($Bo \sim O(10^{-3})$). Among many notable contributions to the field, outlined in a recent review \cite{Yu2023BubbleReview}, the interrelation between the nature of the capillary waves and the ejection of the Worthington jet is now well-understood as a result of the works of Refs. \cite{Gordillo2019CapillaryJets,Blanco-Rodriguez2021OnBubbles} and collaborators. The theoretical models proposed in these references suggest that the Worthington jet forms due to the system's selection of a dominant capillary wave with wavelength $\lambda^{*}$ during thin film retraction. This wavelength, which scales with $Oh$ as $\lambda^{*} \sim Oh^{1/2}$, corresponds to that of the smallest wavelength not fully attenuated by viscous effects. These studies have also uncovered the constant behaviour of the angular velocity of the dominant capillary wave, $|u_{\theta, lin}| \approx 5u_{c}$, as it propagates across the bubble, while also confirming the negligible influence of $Oh$ on its behaviour up until $Bo \approx 0.7$.

Through extensive experiments and numerical simulations, Refs. \cite{Duchemin2002JetSurface,Deike2018DynamicsBubbles,Gordillo2019CapillaryJets,Ganan-Calvo2017RevisionSpeed,Ghabache2016SizeBursting,Lai2018BubbleProfiles,Brasz2018MinimumBubble,Walls2015JetProduction} (and references cited therein) have contributed to characterising the size and velocity of the first ejected drop and the Worthington jet. Particularly noteworthy is the work of Ref. \cite{Gordillo2019CapillaryJets}, which proposes relationships between $Oh$ and the jet's velocity, $u_{jet}$. These relationships include $u_{jet} \sim Oh^{1/2}$ ($Oh << Oh_{*}$), $u_{jet} \sim (1 - (Oh/Oh_{*})^{1/2})^{-1/2}$ ($Oh \approx Oh_{*}$), and $u_{jet} \sim Oh^{-1}$ ($Oh > Oh_{*}$), where $Oh_{*} \approx 0.043$ denotes the $Oh$ value above which viscous effects preclude the jet from breaking. Refs. \cite{Ganan-Calvo2017RevisionSpeed} and \cite{Deike2018DynamicsBubbles} have proposed scaling laws of the form $\sim Oh^{1/2}(Oh_{*} - Oh)^{-3/4}$ to relate the velocity of the first ejected drop or the jet before breakup, $u_{d1}$, with $Oh$, with the latter reference also proposing an expression to include the effects of $Bo$. In terms of first ejected drop size, $R_{d1}$, numerous experimental and numerical works have confirmed the scaling derived in Ref. \cite{Ganan-Calvo2017RevisionSpeed}, in which $R_{d1}$ scales with $Oh$ as $\sim [(Oh_{*} - Oh)Oh^{-2}]^{5/4}$. 


The structurally complex interfacial phenomena that arise from surface-active material have been the subject of a handful of recent investigations. The experimental and numerical work of Refs. \cite{Pierre2022InfluenceBursting, Vega2024InfluenceBursting,Ji2023SecondarySurface,Neel2022RoleBursting,Constante-Amores2021DynamicsInterface} for $Bo \to 0$ have reported dramatic modifications to all steps leading to jet breakup, revealing intricate relationships between the system's outcomes and surfactant properties due to the comparable nature of surfactant-related time-scales and those of bubble bursting. The most notable of these modifications involves a consistent temporal retardation of bubble collapse with the addition of insoluble surfactant (Sodium DodecylSulfate, SDS). This retardation increases in prominence as the surfactant concentration increases but remains below the critical micelle concentration (CMC) \cite{Pierre2022InfluenceBursting}. Simulations by Ref. \cite{Constante-Amores2021DynamicsInterface} have provided initial insights into the complex role of Marangoni stresses and reduced surface tension on the retardation dynamics. This study reported the development of a toroidal capillary wave carrying a large surfactant concentration across the bubble's domain, leading to a low surface tension bubble collapse (larger local $Oh$) and a fast jet ejection ($u_{jet} \sim Oh^{1/2}$ \cite{Gordillo2019CapillaryJets}) that is subsequently delayed by the action of Marangoni stresses. Other surfactant-induced effects unveiled in the literature relate to a reduction in the number and size of ejected drops below the CMC \cite{Pierre2022InfluenceBursting,Vega2024InfluenceBursting}. The recent experiments and theoretical framework presented in Ref. \cite{Ji2023SecondarySurface} expose the marked influence of surface viscoelasticity and its unequivocal departure from Newtonian surface behaviour in surfactant-laden flows. This study introduces a new bursting regime featuring secondary daughter bubble entrapment during collapse and jetting suppression, which uniquely originate from the action of dilatational elastic stresses rather than Marangoni stresses. A similar jetting suppression phenomenon has been observed in bursting bubble systems in non-Newtonian liquid media and surfactant-free interfaces \cite{Sanjay2021BurstingMedium}.

Despite recent strides in untangling the rich interfacial phenomena that unfold in the presence of surface-active contaminants, a few critical aspects remain elusive. While experimental observations have detected a delay in capillary wave motion, enabling estimations of capillary wave velocity in surfactant-laden systems, temporal information on the evolution of the angular position of the dominant capillary wave, $\theta$, akin to the insights provided by Ref. \cite{Gordillo2019CapillaryJets} for clean interfaces, has not been reported before. As demonstrated throughout this manuscript, uncovering the dependence of $\theta$ on $t$ is paramount to understanding the system's dynamics, representing the inherent nonlinearity and complexity of surfactant flows that are heavily influenced by surfactant parameters. Additionally, and as stated by Refs. \cite{Pierre2022InfluenceBursting,Vega2024InfluenceBursting}, the exact mechanism by which the motion retardation comes about, as well as the unique influence of Marangoni stresses, is still unclear and challenging to access experimentally. The present manuscript builds upon previous work and presents a numerical exploration of the problem of bursting bubbles in surfactant-laden flows of varying solubility and surfactant strength, operating under non-negligible gravitational conditions ($Bo = 0.1 - 5$). To the best of the authors' knowledge, this is the first time a mechanistic rationalisation of the specific role of Marangoni stresses in wave motion delay has been proposed in surfactant-laden bursting bubbles. This manuscript also illustrates the effects of non-vanishing $Bo$ on capillary waves for values ($Bo > 0.7$) not previously reported in the literature, and how gravitational and surfactant effects converge to fully suppress jetting.

The rest of this manuscript is organised as follows: §\ref{sec:methods} introduces the system considered, our numerical approach, and related dimensionless parameters. §\ref{sec:results} presents our results and discussion, focusing first on surfactant-free interfaces at large $Bo$, and then inspecting the striking influence of surfactants. Concluding remarks are provided in §\ref{sec:conclusions}.



\section{Formulation and problem statement}\label{sec:methods}

\subsection{Governing equations and dimensionless numbers}\label{subsec:dimless}

We conduct two-phase direct numerical simulations to investigate the dynamics of an isolated bursting bubble event immersed in a three-dimensional (3-D) Cartesian domain denoted by $\textbf{x}=(x,y,z)$ (refer to figure \ref{fig:sim_conf}(a)). The simulations incorporate surface-active material, and the equations presented in this manuscript are formulated within the framework of a hybrid font-tracking/level-set technique known as the ``Level Contour Reconstruction Method" (LCRM). This method is employed to handle surface tension forces, the interface, and their coupling with surfactant transport. Details of this approach can be found in Refs. \cite{Shin2009ATechniques,Shin2017A,Shin2018ASurfactant}. All variables and physical parameters involved in the system are rendered dimensionless via the scalings expressed by Eq. (\ref{eq:scaling}), which are consistent with previous work on bursting bubbles \cite{Constante-Amores2021DynamicsInterface,Gordillo2019CapillaryJets}: 
\begin{equation}\label{eq:scaling}
\quad \tilde{\mathbf{x}}=\frac{\mathbf{x}}{R_{0}},
\quad \tilde{t}=\frac{t}{t_{c}}, 
\quad \tilde{\textbf{u}}=\frac{\textbf{u}} {u_{c}},
\quad \tilde{p}=\frac{p}{\rho u_{c}^2}, 
\quad \tilde{\mu}=\frac{\mu}{\mu_{l}},
\quad \tilde{\rho}=\frac{\rho}{\rho_{l}}, 
\quad \tilde{\sigma}=\frac{\sigma}{\sigma_s},
\quad \tilde{\Gamma}=\frac{\Gamma}{\Gamma_\infty},
\quad \tilde{C}=\frac{C}{C_{\infty} },
\quad \tilde{C}=\frac{C_s}{C_{\infty} },
\end{equation}
\noindent where $\tilde{t}$, $\tilde{\textbf{u}}$, $\tilde{p}$, $\tilde{\mu}$, $\tilde{\rho}$, and $\tilde{\sigma}$ represent time, velocity vector, pressure, and local viscosity, density, and surface tension, respectively; $\tilde{\Gamma}$, $\tilde{C}$, and $\tilde{C}_{s}$ are surfactant concentration at the interface, bulk, and bulk region in the vicinity of the interface, respectively. Hereinafter, tildes represent dimensionless variables, and the subscripts $l$ and $g$ distinguish between the liquid and gas phases. In the above equations, $R_{0}$ is the bubble size, $u_{c} = \sqrt{\sigma_{s}/(\rho_{l}R_{0})}$ is the velocity time-scale, $t_{c} = R_{0}/u_{c} = \sqrt{\rho_{l}R_{0}^{3}/\sigma_{s}}$ is the capillary time-scale, and $\sigma_{s}$ is the surface tension in a surfactant-free interface. Surfactant-related variables are scaled on the saturation interface surfactant concentration or maximum packing, $\Gamma_{\infty}$, and the initial bulk surfactant concentration, $C_{\infty}$. In what follows, we refer to all variables by their name to refer to their dimensionless version, unless stated otherwise.

The equations governing the simulation, presented herein, are expressed in dimensionless form under the assumptions of Newtonian fluids, immiscible phases, and incompressible flow. The continuity and momentum equations, written within a one-fluid formulation, are provided in Eq. (\ref{eq:mass_cons})-(\ref{eq:rho_mu}),
\begin{equation}
 \nabla \cdot \tilde{\textbf{u}}=0,
  \label{eq:mass_cons}
\end{equation}
\begin{equation}
\tilde{\rho} \left(\frac{\partial \tilde{\textbf{u}}}{\partial \tilde{t}}+\tilde{\textbf{u}} \cdot\nabla \tilde{\textbf{u}}\right) + \nabla \tilde{p}  =  -Bo \textbf{i}_z + Oh ~ \nabla\cdot  \left [ \tilde{\mu} (\nabla \tilde{\textbf{u}} +\nabla \tilde{\textbf{u}}^T) \right ] +
\int_{\tilde{A}\tilde{(t)}} 
(\tilde{\sigma} \tilde{\kappa} \textbf{n} +   \nabla_s  \tilde{\sigma})  \delta \left(\tilde{\textbf{x}} - \tilde{\textbf{x}}_{_f}  \right)\mbox{d}\tilde{A},
\label{eq:momentum_cons}
\end{equation}
\begin{equation} \label{eq:rho_mu}
    \left.\begin{array}{c}
    \tilde{\rho}\left( \textbf{x},t\right) = \rho_{g}/\rho_{l} + (1 - \rho_{g}/\rho_{l})\mathcal{H}(\tilde{\textbf{x}}, \tilde{t}),\\
    \tilde{\mu}\left( \textbf{x},t\right)=\mu_{g}/\mu_{l} + (1 - \mu_{g}/\mu_{l})\mathcal{H}(\tilde{\textbf{x}}, \tilde{t}),\\
    \end{array}\right.
\end{equation}
\noindent where $\mathcal{H}$ is a smoothed Heaviside function that adopts the value of zero in the gas phase and unity in the liquid phase. This function is computed directly from the interface and solved numerically with a smooth transition across three or four cells \cite{Shin2009ATechniques,Shin2017A}. As introduced in §\ref{sec:introduction}, $Bo$ denotes the Bond number, and $Oh$ is the Ohnesorge number (see Eq. (\ref{eq:dimless_numbers}) for their definitions). In Eq. (\ref{eq:momentum_cons}), the last two terms on the right-hand side (RHS) account for the normal and tangential components of the surface tension forces, $\tilde{\textbf{F}}_{\sigma}$. The normal component arises from the average localised surface tension, while the tangential component, or Marangoni stresses, acts on the interface due to surfactant-induced surface tension variations across the interface. In the definitions, $\tilde{\kappa}$ is twice the mean interface curvature, $\textbf{n}$ represents a unit vector normal to the interface pointing towards the liquid phase, $\tilde{A}(\tilde{t})$ denotes the time-dependent interface area, and $\delta (\tilde{\textbf{x}} - \tilde{\textbf{x}}_{f})$ is a Dirac delta function. This delta function vanishes everywhere except for the interface (located at $\tilde{\textbf{x}} = \tilde{\textbf{x}}{_f}$); $\nabla_s = (I - \textbf{n}\textbf{n}) \cdot \nabla$ signifies the surface gradient operator.

Surfactant transport is resolved in both the liquid bulk phase and at the interface via the non-passive convection-diffusion expressions of Eq. (\ref{eq:gamma_cons}) - (\ref{eq:C_cons}):
 \begin{equation}\label{eq:gamma_cons}
 \frac{\partial \tilde{\Gamma}}{\partial \tilde{t}}+\nabla_s \cdot (\tilde{\Gamma}\tilde{\textbf{u}}_{\text{t}})=\frac{1}{Pe_s} \nabla^2_s \tilde{\Gamma}+ Bi \left ( k  \tilde{C_s} (1-\tilde{\Gamma})- \tilde{\Gamma}  \right ),
 \end{equation}
\begin{equation} \label{eq:C_cons}
\frac{\partial \tilde{C}} {\partial \tilde{t}}+\tilde{\textbf{u}}\cdot \nabla \tilde{C}= \frac{1}{Pe_c} \nabla^2 \tilde{C},
\end{equation}
\begin{equation}\label{eq:bulksource_nd}
\textbf{n}\cdot\nabla \tilde{C} |_{interface}=-Pe_c Da Bi \left ( k  \tilde{C_s} (1-\tilde{\Gamma})- \tilde{\Gamma}  \right ),
\end{equation}
\noindent where $\tilde{\textbf{u}}_{\text{t}} = (\tilde{\textbf{u}}_{\text{s}} \cdot \textbf{t})\textbf{t}$ is the tangential projection of the surface velocity vector, $\tilde{\textbf{u}}_{\text{s}}$. 
Within this formulation, surfactant is exchanged between the interface and the bulk region immediately adjacent via the source term written in Eq. (\ref{eq:bulksource_nd}).
The dependence of surface tension on variations in $\tilde{\Gamma}$ is computed via the nonlinear Langmuir expression, which reads
\begin{equation}\label{eq:langmuir}
\tilde{\sigma}=\max\left(\epsilon_{\sigma}, 1 + \beta_s \ln{\left(1 -\tilde{\Gamma}\right)}\right),
\end{equation}
\noindent with $\beta_{s}$ representing the surfactant elasticity, or surfactant ``strength", and $\epsilon_{\sigma} = 0.05$. The latter parameter corresponds to a limit placed on $\tilde{\sigma}$ and introduced in the formulation to circumvent the issue of unphysical negative surface tensions predicted by the Langmuir expression in non-dilute systems where surfactant concentration approaches the maximum packing ($\tilde{\Gamma} \to 1$) \cite{Muradoglu2014SimulationsFlow,Shin2018ASurfactant}. The dimensionless groups that govern the system are defined here as
\begin{equation}\label{eq:dimless_numbers}
Bo = \frac{\rho_{l} g R_{0}^{2}}{\sigma_{s}}, ~~~
La = Oh^{-2} = \frac{\rho_{l} \sigma_{s} R_{0}}{\mu_{l}^{2}}, ~~~
Bi = \frac{k_{d} R_{0}}{u_{c}},        ~~~
k = \frac{k_{a} C_{\infty} }{k_{d}}, ~~~
Da = \frac{\Gamma_{\infty}}{R_{0}C_{\infty}}, ~~~
Pe_{s,c} = \frac{u_{c} R_{0}}{D_{s,c}},     ~~~ 
\end{equation}
\noindent where $Bo$ characterises the interplay between gravitational and capillary forces, while $La$ quantifies the relative significance of viscous to capillary forces; The Biot number, $Bi$, serves as a metric for surfactant solubility, comparing the capillary and desorptive, $t_{d} = k_{d}^{-1}$, timescales. The parameter $k$ directly influences the balance between surfactant adsorption, $t_{ad} = (k_{a}C_{\infty})^{-1}$, and desorption time-scales. The Damköhler number, $Da$, provides insight into the initial surfactant saturation, while $Pe_{s,c}$ denotes the Péclet numbers associated with the significance of surfactant diffusive effects on both the interface (modulated by diffusivity $D_{s}$) and the liquid bulk (modulated by diffusivity $D_{c}$) with respect to the characteristic timescale of the system, $t_{c}$. 

In the present study, we compute the dimensionless Marangoni stresses, $\tilde{\tau}_{m}$, by using Eq. (\ref{eq:marangoni})
 \begin{equation}\label{eq:marangoni}
 \tilde{\tau}_{m} \equiv  \nabla_s \tilde{\sigma} \cdot  \textbf{t} =-\frac{\beta_s}{1 -\tilde{\Gamma}} \nabla_s\tilde{\Gamma} \cdot \textbf{t}.
\end{equation}
The variation of $\tilde{\tau}_{m}$ with respect to $\beta_{s}$ positions surfactant strength as a regulator of Marangoni stress magnitude. In our convention, positive values of Marangoni stresses are observed in regions transitioning from higher to lower $\tilde{\Gamma}$.

\subsection{Numerical method}\label{subsec:num_method}

For our calculations, we employ the code BLUE, which is based on the LCRM method. This method and code have been successful in reproducing both experimental and numerical data in a plethora of applications involving complex interfacial flow instabilities with contaminants, including drop coalescence \cite{Constante-Amores2021RoleCoalescence}, ligament breakup \cite{Constante-Amores2020DynamicsNumber}, turbulent jets \cite{Constante-Amores2023DirectInteractions}, microfluidics \cite{Kalli2023EffectApproach}, falling film \cite{Batchvarov2020Three-dimensionalSurfactants}, elongated bubbles \cite{Batchvarov2020EffectNumber,Pico2024DropChannels} and indeed, bursting bubbles for very small Bond number \cite{Constante-Amores2021DynamicsInterface}. We present in this sub-section a brief summary of the most relevant aspects of our method and refer the reader to Refs. \cite{Shin2009ATechniques,Shin2017A,Shin2018ASurfactant} for further details. The field equations presented above are resolved by utilising a finite-difference method with a fixed Eulerian grid. Spatial derivatives of the Eulerian grid fields are discretised through a standard cell-centred scheme for most terms, except for the nonlinear convective terms, which employ a second-order essentially non-oscillatory (ENO) procedure \cite{Shu1989EfficientII} on staggered grid. The viscous term in the momentum equation is discretised using a second-order centred difference scheme. Temporal terms are handled with a second-order Gear method \cite{Wang2006GearSystems}, incorporating implicit time integration for the viscous terms. 

The interface is tracked with a front-tracking method and reconstructed using a level-set approach \cite{Shin2002ModelingConnectivity,Shin2009ATechniques}. The connection between Lagrangian and Eulerian grid is achieved through \citet{Peskin1977NumericalHeart}'s well-known immersed boundary method. The front-tracking utilises a deformable grid with Lagrangian elements discretised using a triangular mesh. These Lagrangian elements are advected by integrating equation $\textrm{d}{\textbf x}_{f}/\textrm{d}t = \textbf{V}_{f}$, where $\textbf{V}_{f}$ represents the interface fluid velocity, which displaces the vertices of the Lagrangian triangular mesh, $\textbf{x}_{f}$. This integration is performed using a second-order Runge-Kutta method.

\subsection{Simulation setup and timescales}\label{subsec:sim_setup}

The computational domain for most simulations consists of a rectangular box of size $12R_{0} \times 12R_{0} \times 6R_{0}$, where the latter dimension corresponds to the domain's height (see figure \ref{fig:sim_conf}(a)). We highlight that this height is smaller than that of previous numerical studies, such as the $15R_{0}$ used in Ref. \cite{Constante-Amores2021DynamicsInterface} or the $8R_{0}$ used in Ref. \cite{Sanjay2021BurstingMedium} as our focus is specifically on the large Bond ($Bo = 0.5 - 5$) regime. In this regime, the system exhibits a significant reduction in the velocity and length of the jet \cite{Deike2018DynamicsBubbles,Berny2020RoleBubbles}. During testing, we found this domain size to be appropriate to render radially- and height-related boundary effects negligible. For the simulations characterised by $Bo = 0.1$, the domain size was increased to $(12R_{0})^{3}$.

The system is initiated with a static equilibrated bubble positioned beneath the liquid-gas interface, shaped according to the Young-Laplace equation. This equation dictates the bubble's morphology based on a static quasi-equilibrium between surface tension and gravity, directly influenced by the choice of $Bo$ \cite{Lhuissier2012BurstingAerosols,Zhang2020NumericalAlgorithms}. A cross-section of the initial bubble shapes for the four tested $Bo$ values is depicted in figure \ref{fig:sim_conf}(b). Notice that, as $Bo$ increases, there is a clear departure from sphericity, accompanied by reduced immersion in the liquid bulk. In adherence to the common approach in numerical investigations of bursting bubbles, even at $Bo > 1$ \cite{Walls2015JetProduction}, the simulations commence at the point when the liquid cap completes its draining process. This simplification, enabling the exclusion of the liquid cap, is justified by the markedly larger Taylor-Culick velocity associated with cap retraction time, $u_{TC}$, in comparison to the system's main phenomena. Under our conditions, $u_{TC}$ is larger than our characteristic velocity scale $u_{c}$ by three to two orders of magnitude. We nonetheless discuss the implications of neglecting the cap dynamics at large $Bo$ and the caveats of our results at the end of §\ref{subsec:results_clean} and Appendix \ref{subsec:cap}.

We employ a structured uniform grid of resolution $(768)^{2} \times 384$ ($(768)^{3}$ for $Bo = 0.1$), ensuring converged and mesh-independent results during testing for both the present study and our prior work on bursting bubbles \cite{Constante-Amores2021DynamicsInterface}. This grid resolution proves crucial, particularly at the instants of jet ejection and breakup, where large velocities are involved. For a comprehensive validation of our code, extending beyond our prior work on bursting bubbles \cite{Constante-Amores2021DynamicsInterface}, please refer to Appendix \ref{subsec:validation}. We seek solutions to the governing equations subject to Neumann boundary conditions for the lateral and bottom boundaries in all variables, except for pressure, for which no-slip conditions are utilised. Surfactant material is initialised with an equilibrium condition between the bulk and interface. By equating the last term on the RHS of Eq. (\ref{eq:gamma_cons}) to zero and considering $\tilde{C} = 1 = \tilde{C}_{s}$, the initial interface surfactant concentration is set to $\tilde{\Gamma}(\textbf{x}, t = 0) = \tilde{\Gamma}_{0} = k/(k+1) = 0.55$; this concentration is kept constant in all simulations.

We examine the impact of surfactant solubility within the range $Bi = 0 - 10$ and surfactant strength within the range $\beta_{s} = 0.05 - 1$. As implied in Eq. (\ref{eq:dimless_numbers}), our formulation specifies that $Bi = 0$ denotes an insoluble surfactant (i.e., no surfactant desorption), while the limit $Bi \to \infty$ corresponds to a surfactant-free interface. The additional dimensionless parameters associated with surfactant remain constant at $k = 1.22$, $Pe_{s,c} = 100$, and $Da = 1$. Along with the above parameters, we vary both $Bo$ and $La$ in the ranges $0.1 < Bo < 5$ and $10^{4}< La < 10^{6}$ ($10^{-2} > Oh > 10^{-3}$). These $La$ conditions are well above the ``optimal" value of $La \approx 1000$, where the system approaches a singularity at the instant of collapse and the maximum (minimum) jet velocity (drop size) is attained for $Bo \to 0$ \cite{Berny2020RoleBubbles,Gordillo2019CapillaryJets}.

Our selection of dimensionless parameter values is informed by the characteristics of commonly employed commercial surfactants in bursting bubble applications, such as sodium dodecyl sulfate (SDS) and Triton X-100 (TX100) \cite{Pierre2022InfluenceBursting,Vega2024InfluenceBursting}. These surfactants exhibit diffusivities in the range of $D_{s,c} = \textit{O}(10^{-10} - 10^{-9})$ $\rm{m^{2}s^{-1}}$, resulting in corresponding Péclet numbers of $Pe_{s,c} = \textit{O}(10^{3} - 10^{6})$. This indicates that surfactant transport and rearrangement events across the interface are unlikely to be dominated by interfacial surfactant diffusion. Our chosen conditions also ensure the significant involvement of Marangoni stresses in our system, given the comparable nature of the Marangoni time-scale, $t_{m} = \mu_{l}R_{0}/(\Delta \sigma) = \textit{O}(10^{-2})$ $\textrm{s}$, with those associated with sorption kinetics, $\textit{O}(10^{-2}) - \textit{O}(10^{-1})$ $\textrm{s}$ \cite{Mayer2012Landau-LevichPhenomena}. Furthermore, we operate at sufficiently low bulk surfactant concentrations to neglect the micelle formation and its potential effects on the system.


\section{Results and discussion}\label{sec:results}

\subsection{Uncontaminated interfaces: buoyancy effects on capillary wave velocity and Worthington jet}\label{subsec:results_clean}

\begin{figure}
\includegraphics[width=1.02\linewidth]{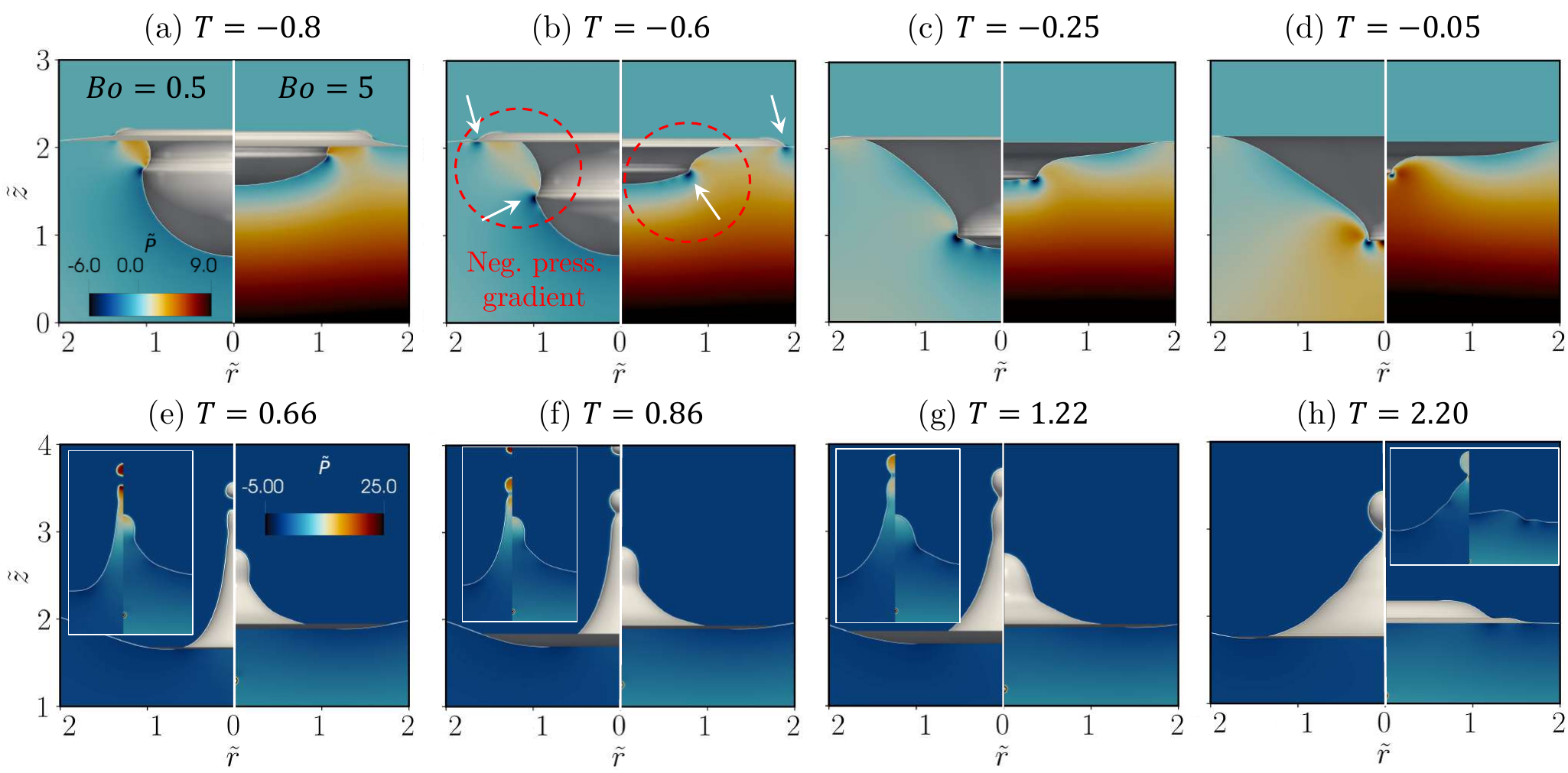}
  \caption{Bubble bursting dynamics in surfactant-free interfaces: comparison of pressure fields and interface shape for a moderate $Bo = 0.5$ on the left half and a large $Bo = 5$ on the right half: (a)-(d) pre-collapse dynamics; (e)-(h) Worthington jet motion and drop detachment. The insets in (e)-(h) focus on the emerging jet's pressure fields. For all cases in this figure, $Oh = 3.21 \times 10^{-3}$ ($La = 9.7\times10^{4}$).}
  \label{fig:press_clean}
\end{figure}

Let us begin our discussion by analysing the influence of gravitational effects in surfactant-free interfaces. Figure \ref{fig:press_clean} offers a qualitative comparison of the system's behaviour pre- (a)-(d) and post- (e)-(h) bubble collapse for $Bo = 0.5$ and $Bo = 5$ in a representative case at $Oh = 3.21\times10^{-3}$. In order to make temporal one-to-one comparisons between different cases and bursting stages, we define a comparative time, $T = (\tilde{t} - \tilde{t}_{0})/\tilde{t}_{0}$, which takes the bubble collapse time, $t_{0}$, as a reference and compares the system's dynamics at the same temporal distance from collapse. Thus, $T < 0$ ($T > 0$) designates the pre-collapse (post-collapse, or jetting) stages. The pressure fields depicted in the figures are presented in relation to the pressure at the pool-atmosphere interface. As detailed in §\ref{sec:introduction}, the process of liquid cap retraction concludes in the outset of capillary wave motion away and down the cavity. The latter set of waves ultimately converges at the bubble's bottom, giving rise to the Worthington jet. The dominant capillary waves (i.e., those characterised by the largest curvature \cite{Gordillo2019CapillaryJets}) are clearly seen at the low-pressure locations marked by white arrows.

During the early and middle bursting stages ($T < -0.25$), the high-pressure regions immediately upstream of the bubble's dominant wave create a negative pressure gradient, directing the flow downward in the $z$-direction and facilitating capillary wave motion towards the bottom of the cavity, as depicted in red in the figure. In the case of $Bo = 5$, this negative pressure gradient competes with the upward pressure gradient between the high-pressure liquid bulk and the low-pressure region beneath the bubble's bottom. The discussion of subsequent figures will elucidate how this competition between pressure gradients is accountable for delaying capillary wave motion, resulting in a lower average motion velocity above $Bo = 1$. Furthermore, we will explore how these retarding gravitational effects do not impact the linearity of the motion, in contrast to the effects induced by surfactant.

The post-collapse phenomena of figure \ref{fig:press_clean}(e)-(h) underscore the anticipated stabilising influence of gravity on capillary breakup and jet growth, resulting in complete avoidance of pinch-off and notably smaller jets for $Bo = 5$ across all tested $La$ values (refer to Appendix \ref{subsec:validation} for a compilation of the effects of $La$ and $Bo$ on jet length and velocity at the instant before the first pinch-off, as well as the size of the first ejected drop). As is evident from the interfacial shapes in figure \ref{fig:press_clean}, the deceleration of the jet due to gravitational forces continues to shape the jet's dynamics following the initial pinch-off event for $Bo = 0.5$. Notably, we observe that the maximum dimensionless jet length achieved, irrespective of pinch-off occurrences, exhibits a relationship with $Bo$ given approximately by $(L_{jet}/l_{vis})_{\textrm{Max}} \sim Bo^{-0.36}$ across all three considered $La$ values, in contrast to the moment of the first pinch-off where $(L_{jet}/l_{vis}) \sim Bo^{-0.16}$. Here, $l_{vis}$ denotes the viscous length scale, defined in Eq. (\ref{eq:scales_viscosity}). This inverse relationship between jet length and $Bo$ is in agreement with the observations of Ref. \cite{Walls2015JetProduction}. We highlight that the threshold for jet drop production in our simulations also aligns completely with the $Oh-Bo$ maps indicating production/no production from the same reference (see figure 4 in that reference).

\begin{figure}
\includegraphics[width=0.95\linewidth]{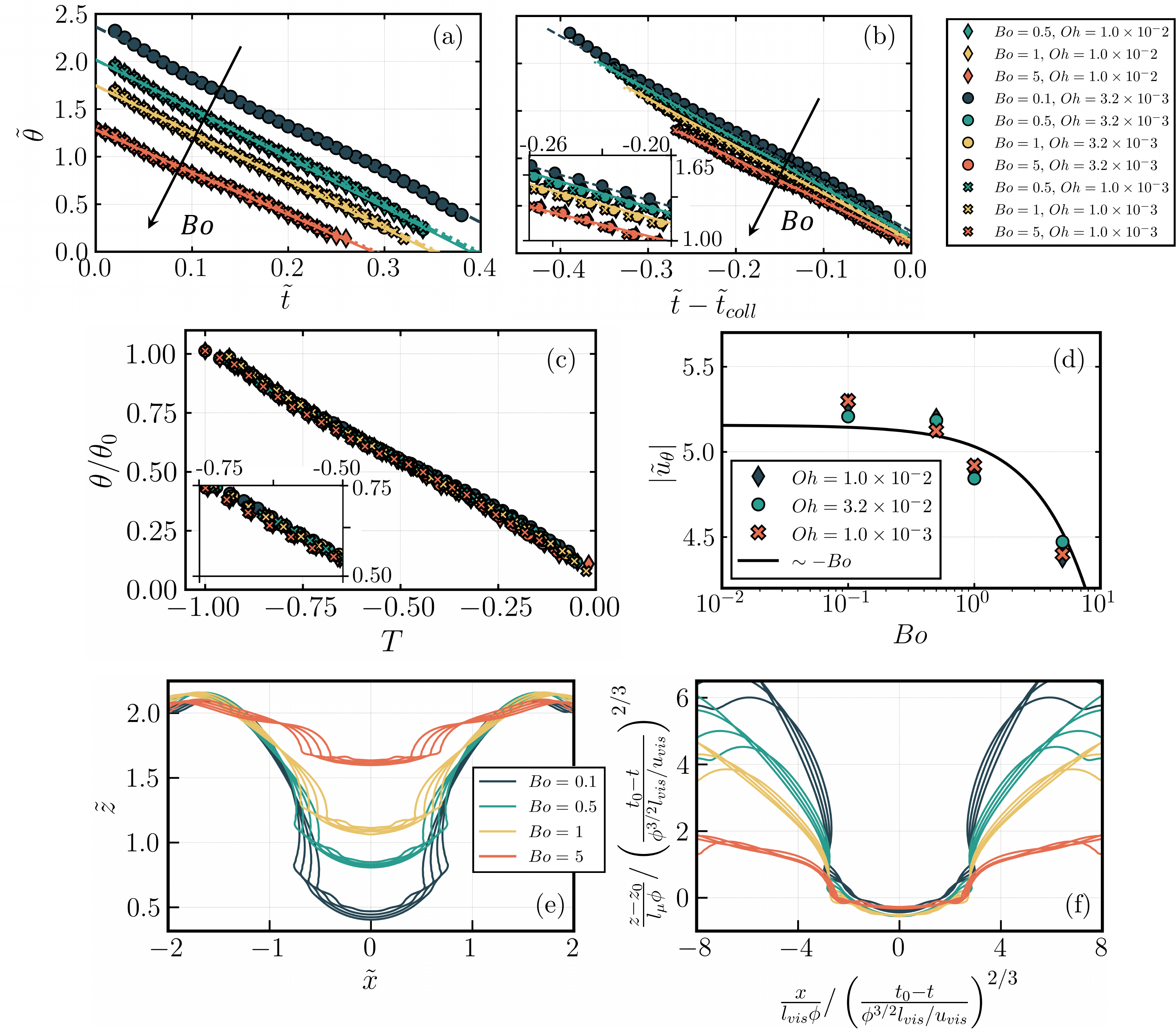}
  \caption{Dynamics of capillary wave velocity and interfacial shape in surfactant-free interfaces: effect of $Bo$ and $Oh$; (a) temporal evolution of $\tilde{\theta}$; (b) comparison of the slopes of the functions $\tilde{\theta}$ vs. $\tilde{t}$, showcasing the minimal influence of $Oh$ and the slight influence of $Bo$ above $Bo = 1$; (c) data scaled by $\theta_{0}$ and $\tilde{t}_{0}$, showcasing the constant nature of the capillary wave velocity, even at the largest $Bo$ tested, $Bo = 5$; (d) compilation of $|\tilde{u}_{\theta}|$ vs. $Bo$ data wherein the markers represent our simulation data (taken every $\Delta \tilde{t} = 0.01$) and the line corresponds to a linear fit of $-0.126 Bo$ with $R_{pow}^{2} = 0.9833$. ; (e)-(f) interfacial shapes for varying $Bo$ across four different times in the range $t_{0} - t \approx 6t_{pre} - 10t_{pre}$ in the standard Cartesian space and in the rescaled spaced proposed in \cite{Lai2018BubbleProfiles}; $z_{0}$ represents the point in the vertical coordinate where jet ejection occurred. For these cases, $Oh = 3.21 \times 10^{-3}$.}
  \label{fig:vel_wave_clean}
\end{figure}

In figure \ref{fig:vel_wave_clean}, we plot our measurements of angular location of the dominant (or ``strongest", as referred to in other works \cite{Sanjay2021BurstingMedium}) capillary wave, $\tilde{\theta}$, as it transverses down the bubble's domain. This dominant wave corresponds to the sharpest (of lowest wavelength) wave not fully dampened by viscous dissipation \cite{Gordillo2023TheoryJets,Gordillo2019CapillaryJets}. Our $\tilde{\theta}$ numerical observations align with the methodology outlined by \citet{Gordillo2019CapillaryJets}, where the angle for each $Bo$ is measured from the bubble's initial centre-point, as illustrated in figure \ref{fig:sim_conf}(b)-(c). In Part (a) of figure \ref{fig:vel_wave_clean}, we observe a nearly complete marker superposition in the $\tilde{\theta}$ $vs.$ $\tilde{t}$ data for all $Oh$ tested under the same $Bo$. This alignment not only reaffirms prior findings indicating a negligible influence of $Oh$ on capillary wave velocity up to $Bo = 0.7$, as demonstrated by previous researchers (see supplementary information in Ref. \cite{Blanco-Rodriguez2021OnBubbles}), but also extends this observation to higher Bond numbers ($Bo < 5$). It is essential to note that the downward shift of the data for increasing $Bo$ values is attributed to variations in the initial angle, $\tilde{\theta}_{0}$, which itself results from differences in the initial submersion degrees of the bubbles for different $Bo$.

Figure \ref{fig:vel_wave_clean}(b) presents a comparison of the slopes of the functions $\tilde{\theta}$ $vs.$ $\tilde{t}$ by shifting the $x$-axis relative to $\tilde{t}_{0}$. It is noteworthy that, akin to prior studies utilising other numerical methods, the emergence of high velocities and pressures as collapse is approached impedes the unambiguous tracking of $\tilde{\theta}$ as $\tilde{t} \to \tilde{t}_{0}$. Consequently, our measurements do not precisely terminate at $\tilde{\theta} = 0$. The calculated slope values from our available data (normalised by $u_{c}$), denoted as $\tilde{u}_{\theta} = u_{c}^{-1}(\theta_{f} - \theta_{0}) / t_{f}$, are compiled in figure \ref{fig:vel_wave_clean}(d). Inspection of these values reveals a transition from $|u_{\theta}| \approx 5$ (fully consistent with Refs. \cite{Blanco-Rodriguez2021OnBubbles,Gordillo2019CapillaryJets}) at $Bo = 0.1, 0.5, 1$ to $|\tilde{u}_{\theta}| \approx 4.35$ at $Bo = 5$. Although small, this velocity decrease is found to persist for all $Oh$ tested, suggesting an approximate relation of the form $|\tilde{u}_{\theta}| \sim -Bo$, as seen by the curve displayed in the figure for reference. Figure \ref{fig:vel_wave_clean}(c), along with the $R^{2}$ values of a linear fit (ranging from 0.998 to 0.999 in all cases) clearly illustrate the constant nature of the capillary wave velocity in surfactant-free interfaces, even at large $Bo$. The linearity of $\tilde{\theta}$ $vs.$ $\tilde{t}$ enables the collapse of all data points using the comparative time, $T = (\tilde{t} - 
\tilde{t}_{0}) / \tilde{t}_{0}$, and the initial location of the wave capillary wave.

This $\tilde{\theta}$ collapse can be related to the well-known interfacial shape self-similar behaviour featured in uncontaminated interfaces. This self-similarity has been shown to underpin the dynamics of collapsing structures that lead to jetting, such as bursting bubbles and interfacial waves \cite{Lai2018BubbleProfiles,Duchemin2002JetSurface, Ghabache2014OnEjection, Zeff2000SingularitySurface}. An inviscid and incompressible balance of inertia and capillarity in scenarios neglecting gravity leads to an interface shape scaling that obeys $h(r,t) \sim |t_{0} - t|^{2/3}$. The work of Ref. \cite{Lai2018BubbleProfiles} has demonstrated that this scaling is valid before and after bubble collapse in a specific range of temporal distances from collapse. During the pre-collapse period, this range corresponds to $t_{0} - t \approx 6t_{pre} - 10t_{pre}$, where $t_{pre} = \phi^{3/2}l_{vis}/u_{vis}$ is the characteristic time of the capillary wave; $\phi$, $l_{vis}$, and $u_{vis}$ are defined in Eq. \ref{eq:scales_viscosity}. As $t_{0} - t \to 0$ and the curvature inversion that leads to the nascent jet takes place, this self-similarity is disrupted.

Based on the above, we plot in figure \ref{fig:vel_wave_clean}(e)-(f) our computed interfacial shapes for different $Bo$ and for $Oh = 3.21 \times 10^{-3}$ ($La = 97000$) in the standard Cartesian space and in the rescaled space proposed in Ref. \cite{Lai2018BubbleProfiles}. Lines of the same colour correspond to the same $Bo$ at four different times falling approximately within the $t_{0} - t \approx 6t_{pre} - 10t_{pre}$ window. We observe reasonable interfacial collapse in the region surrounding the bottom of the cavity, both across varying times for individual cases and between different $Bo$ values. Although reasonable, the fact that the rescaled profiles do not exactly fall within a single universal line is a testament of the non-negligible gravitational effects present at large $Bo$. We also draw attention to the fact that this scaling was originally reported to be valid only for $La \approx 1000 - 5000$ during pre-collapse, further explaining the incomplete collapse of the curves.

\begin{figure}
\includegraphics[width=0.85\linewidth]{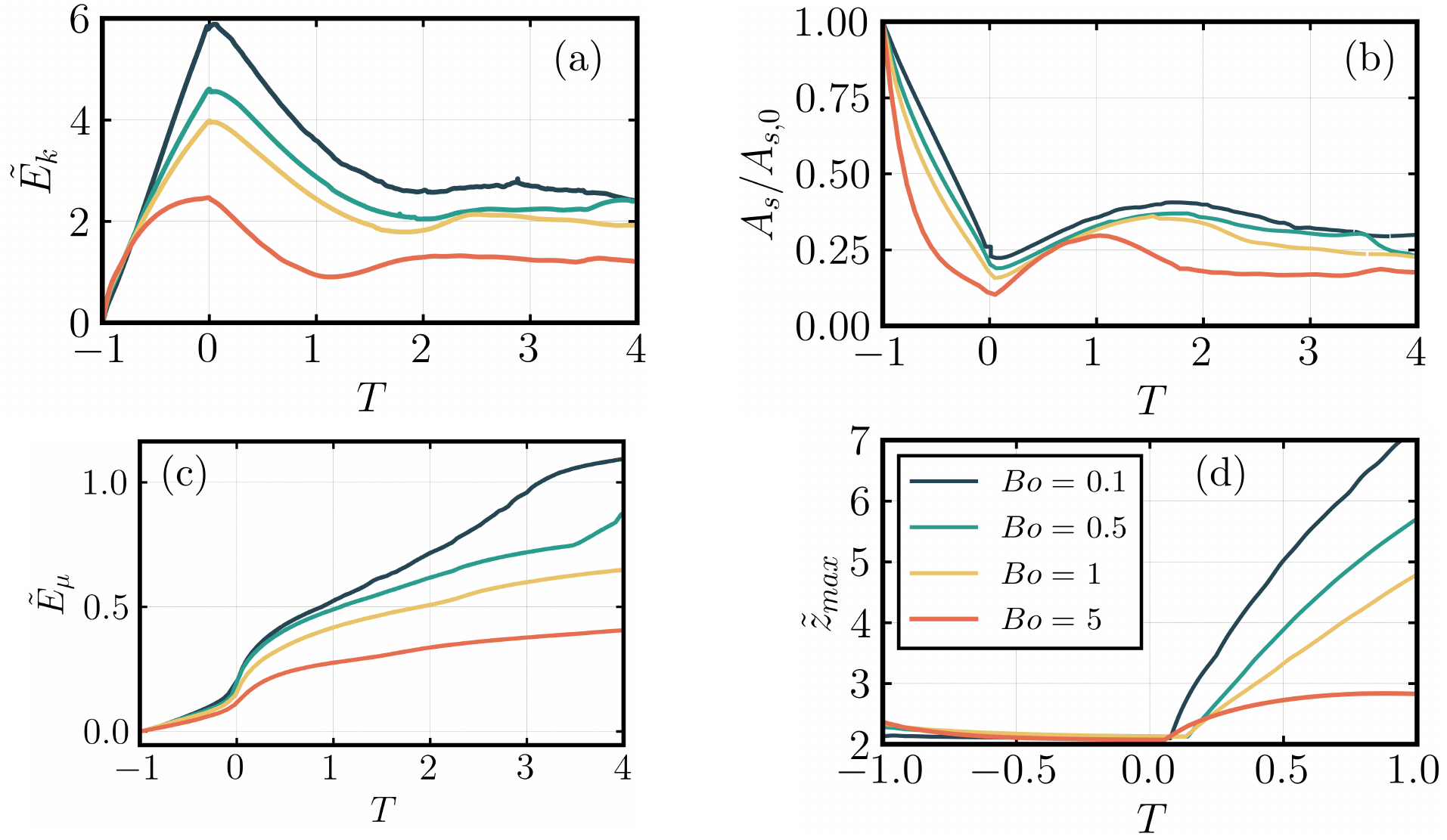}
  \caption{Temporal evolution of kinetic energy, interfacial area, viscous dissipation energy, and maximal vertical location of the interface in surfactant-free cases (i.e., vertical position of the Worthington jet/released drops for $(\tilde{t} - \tilde{t}_{0})/\tilde{t}_{0} > 0$). For all cases in this figure, $Oh = 3.21 \times 10^{-3}$.}
  \label{fig_energy_clean}
\end{figure}

%
%
The influence discussed previously of gravity on both the dominant capillary wave's motion and jet growth/breakup is further illustrated in figure \ref{fig_energy_clean}. This figure presents the temporal evolution of kinetic energy, denoted as $E_{k} = \int_{V}(\rho \textbf{u}^{2}) \, \textrm{d}V$, interfacial area normalised by the initial value, $A_{s,0}$, viscous dissipation energy, $E_{\mu} = \int_{t} (\epsilon_{\mu}) \textrm{d}t$ \cite{Ramirez-Soto2020LiftingOne}, and maximal vertical interface location, $\tilde{z}_{\textrm{max}}$, for various $Bo$; $\epsilon_{\mu}$ is calculated as $\epsilon_{\mu} = \int_{V} (2\mu |\textbf{D}|^{2}) \textrm{d}V$, where $|\textbf{D}|$ is the second invariant of the deformation tensor. Here, $E_{k}$ and $E_{\mu}$ are normalised by $\sigma_{s} R^{2}_{0}$, and we take a flat liquid pool as a reference state to calculate $A_{s}$. The notably faster interfacial energy loss with respect to the system's initial energy (i.e., $A_{s,0}$) for larger $Bo$, combined with their lower energy dissipated by viscous forces at the same $Oh$, and their lower maximal and asymptotic $\tilde{E}_{k}$ values, suggest a heightened presence of gravitational potential energy (with the liquid pool-atmosphere interface as the reference position), influencing the system and retarding motion ($E_{p} \sim Bo$). As previously elucidated, these effects are present during the rise of the Worthington jet (see figure \ref{fig_energy_clean}(d)), bringing about significantly shorter jets and entirely precluding pinch-off for $Bo = 5$.

Before transitioning into the effects of surfactants, it is imperative to discuss a few implications of our computational setup. Specifically, we must address the ramifications of neglecting the dynamics of the retracting liquid cap at large $Bo$. While this approach is common in computational studies, even up to $Bo \approx 5$ \cite{Walls2015JetProduction}, several experimental works have hinted at the emergence of markedly different jets as $Bo$ increases (see, for instance, figure 1 in Ref. \cite{Walls2015JetProduction} or figure 6 in Ref. \cite{Vega2024InfluenceBursting}). As dictated by the Young-Laplace expression, instances characterised by larger $Bo$ entail reduced submersion of the bubble into the liquid pool, leading to a liquid film with much greater surface area exposed to the atmosphere. Consequently, the likelihood of the first cap hole appearing precisely at the flow origin (i.e., at $\tilde{r} = 0$ or $\tilde{x} = 6$ in radial and Cartesian coordinates, respectively), from which fully symmetric cap retraction, capillary wave motion, and jet ejection phenomena unfold, is greatly diminished. The appearance of a liquid cap hole shifted from the axis centre triggers an imbalanced capillary wave motion and a wave convergence point shifted away from $\tilde{x} = 6$, which, in turn, leads to the ejection of non-symmetrical jets. A schematic illustrating this mechanism can be found in Appendix \ref{subsec:cap} (figure \ref{fig_oblique_jet}). The nucleated hole at $\tilde{x} > 6$ elicits a temporal delay in the dynamics at $\tilde{x} < 6$, approximated by the cap retraction time, $t_{TC} \sim E / u_{TC} =  (R^{3}_{0} / l^{3/2}_{c})((\rho_{l})/(2\sigma_{s}\omega))^{1/2}$, where $E \sim R^{2}_{0} / l_{c}$ represents the liquid cap's extent, $l_{c} = \sqrt{2\sigma_{s}/(\rho_{l}g)}$ is the capillary length scale, and $\omega = 10^{3}$ \cite{Villermaux2022BubblesMysteries,Yu2023BubbleReview}. Thus, capillary waves travelling at the $\tilde{x} > 6$ side at a velocity equal to $\approx$ 4 to 5$u_{c}$ meet their counterparts not at $\tilde{x} = 6$, but at $\tilde{x} < 6$, leading to complex wave interactions and non-symmetric jet formation. As this retardation time increases with gravity and $R_{0}$, one can envision a scenario where it becomes comparable to the characteristic time of the problem, $t_{c}$. A comprehensive investigation into this phenomenon of non-symmetric jets and unbalanced capillary wave motion is deferred to future studies.

\subsection{Contaminated interfaces: phenomenology of bubble collapse}\label{sec_surf_wave}

\subsubsection{Surfactant-induced wave retardation and break of linearity}\label{subsec_surf_effect_Bo}

\begin{figure}
\includegraphics[width=1.02\linewidth]{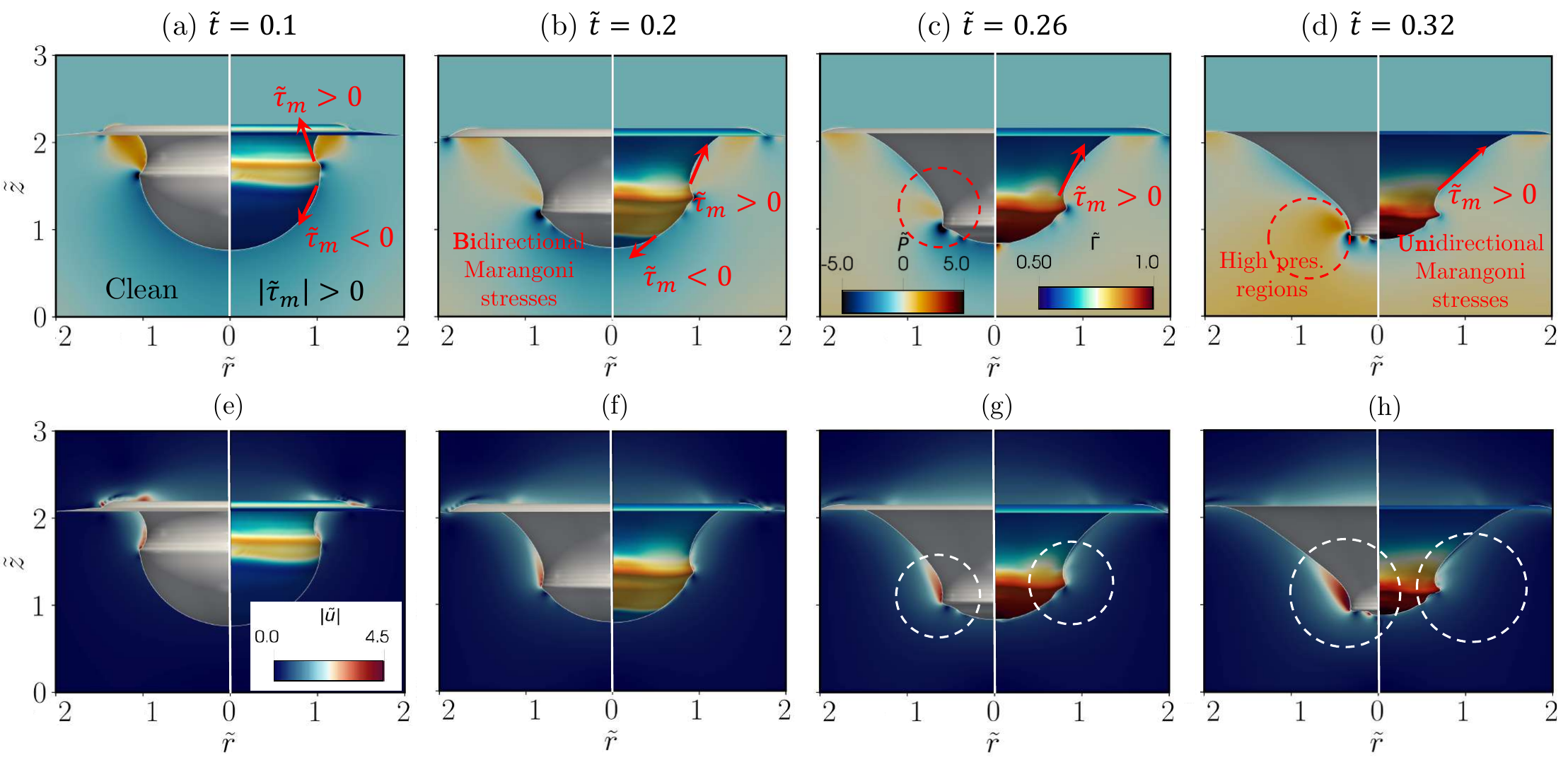}
  \caption{Surfactant-driven capillary wave motion delay across the bubble domain: comparison of 
  temporal evolution of pre-collapse bubble for surfactant-free interfaces on the left half and insoluble surfactant on the right half; (a)-(d) pressure fields; (e)-(h) velocity magnitude fields. For all cases in this figure, $Bo = 0.5$, $Oh = 3.21 \times 10^{-3}$}
  \label{fig_press_vel_surf}
\end{figure}

We now direct our focus towards the intriguing and notably distinct dynamics that emerge with the introduction of surface-active contaminants into the system. In figure \ref{fig_press_vel_surf}, we present a qualitative comparison of wave evolution for two scenarios: a clean case (left half, denoted as ``Clean") and a case with insoluble surfactant characterised by $\beta_{s} = 0.3$ (right half, denoted as ``$|\tilde{\tau}_{m}| > 0$", whose interface is coloured by $\tilde{\Gamma}$), $Bo = 0.5$, and $Oh = 3.21 \times 10^{-3}$. The values for the remaining surfactant-related dimensionless groups are set out in §\ref{subsec:sim_setup}. Each column in the figure corresponds to a distinct dimensionless time, and the background contour plots depict pressure (panels a-d) and velocity magnitude (panels e-h). It is worth noting that we opted not to utilise our comparative time, $T$, in this figure, since both scenarios share the same $Bo$ and our specific focus lies in examining the pre-collapse dynamics from the beginning of the simulation. 

A first inspection of figure \ref{fig_press_vel_surf} reveals a conspicuous delay in dominant capillary wave motion, whose effects are accentuated in time, as will be demonstrated in the upcoming discussions. This phenomenon has previously been reported in a limited number of experimental and numerical studies for various contaminants, including those with surface-active properties \cite{Pierre2022InfluenceBursting,Vega2024InfluenceBursting,Ji2022Water-to-AirDrops,Ji2023SecondarySurface}. Reports of surfactant-driven delays in bubble collapse time have ranged from 35 to 70\% (depending on initial surfactant concentration), which, as we shall see, agree well with our numerical results.
The qualitative similarities between the surfactant-laden and surfactant-free pressure contour plots in figure \ref{fig_press_vel_surf} (discernible dominant capillary wave travelling down and away from the cavity, negative pressure gradient in the $z$-direction, and capillary wave convergence) serve as an indication of the fact that surfactant-induced retardation effects are not sufficiently strong to completely prevent wave focusing and jet formation for the present set of parameters values. This contrasts with observations in non-Newtonian liquids at high plastocapillary numbers \cite{Sanjay2021BurstingMedium}, where capillary wave focusing motion is fully dampened, and the cavity remains open.

\begin{figure}
\includegraphics[width = 1.02\linewidth]{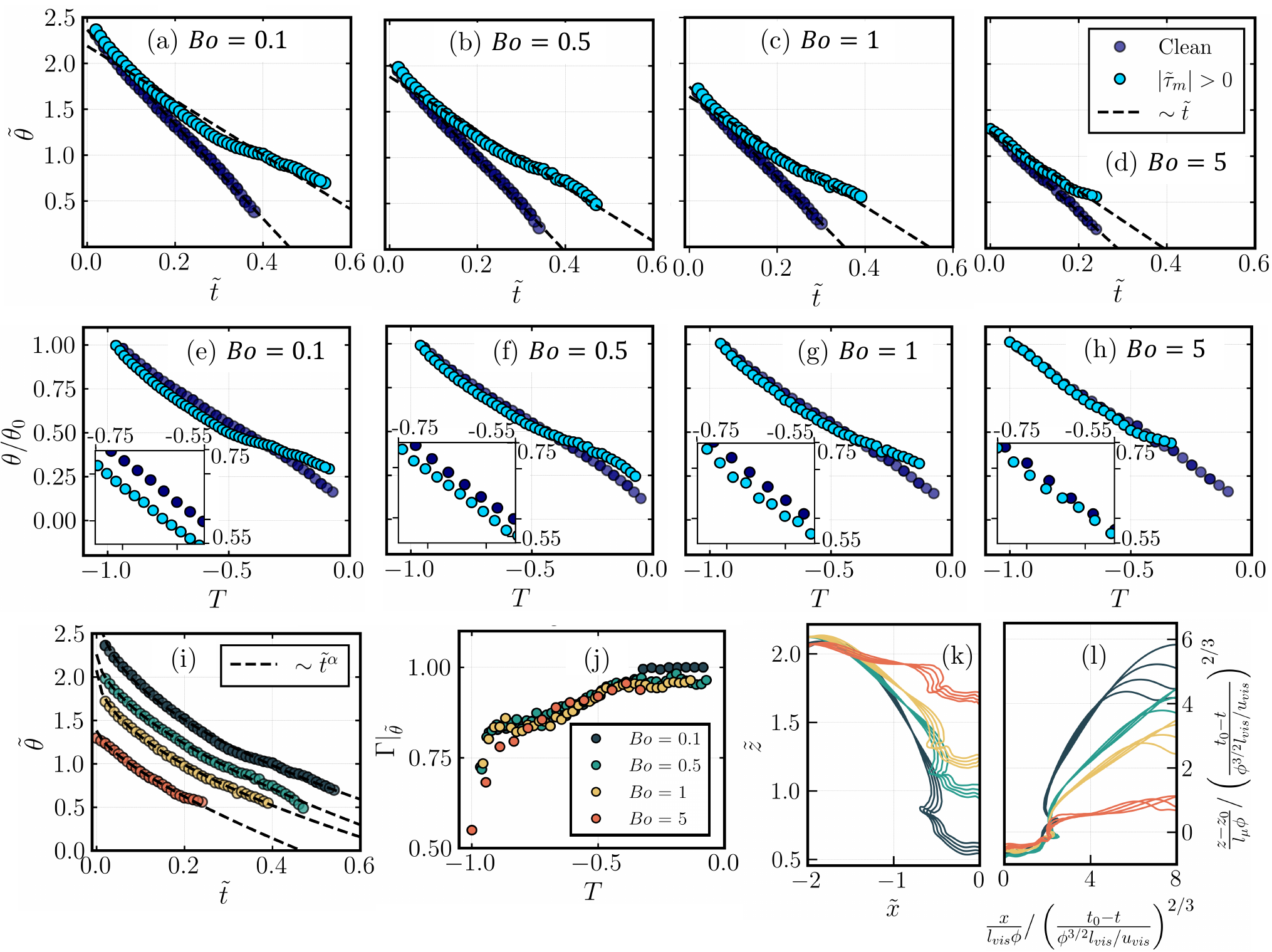}
  \caption{Effect of insoluble surfactant on the dynamics of the capillary wave velocity: (a)-(d) temporal evolution of $\tilde{\theta}$; (e)-(h) data scaled by $\theta_{0}$ and $\tilde{t}_{0}$, showcasing the break of linearity and non-collapsed solutions of contaminated interfaces vs. surfactant-free interfaces; (i) direct comparison of capillary wave dynamics in surfactant-laden cases for varying $Bo$: the markers represent our simulation data (taken every $\Delta \tilde{t} = 0.01$), the lines correspond to linear or power-law fits of the data, and the values of the exponent $\alpha$ for each $Bo$ are found in table \ref{tab:wave_vel_surf}; (j) interface surfactant concentration evaluated at the location of the dominant capillary wave; (k)-(l) interfacial shapes for varying $Bo$ across four different times in the range $t_{0} - t \approx 6t_{pre} - 10t_{pre}$ in the standard Cartesian space and in the rescaled spaced proposed in \cite{Lai2018BubbleProfiles}; $z_{0}$ represents the point in the vertical coordinate where jet ejection occurred. For all the cases in this figure, $Oh = 3.21 \times 10^{-3}$ and $\beta_{s} = 0.3$.}
  \label{fig_vel_wave_surf}
\end{figure}

In the case of surfactants, the initially imposed homogeneity of interface surfactant concentration ($\tilde{\Gamma}_{0} = 0.55$) is promptly destabilised by the substantial capillary pressure stemming from the high-curvature ($|\tilde{\kappa|} \sim O(10)$ at $Oh = 3.21 \times 10^{-3}$) region connecting the bubble and the liquid pool-atmosphere interface (see highlighted region in figure \ref{fig:sim_conf}). As the dominant wave traverses down the bubble, propelled by the negative pressure gradient, a high $\tilde{\Gamma}$ toroidal structure is formed in the vicinity of the dominant wave, from which Marangoni stresses arise (refer to figure \ref{fig_vel_wave_surf}(j) for the temporal evolution of $\tilde{\Gamma}$ evaluated at the location of the dominant capillary wave location, denoted as $\tilde{\Gamma}|_{\tilde{\theta
}}$). The action of these Marangoni stresses, discussed in more extensively in §\ref{subsec_effect_Bi}, ultimately brings about flow retardation and consistently decreases the velocity of the capillary wave, as highlighted in figure \ref{fig_press_vel_surf}(e)-(h).

In figure \ref{fig_vel_wave_surf}(a)-(d), we showcase the temporal evolution of dominant capillary wave angular location obtained from our insoluble surfactant simulations across varying $Bo$ values, juxtaposed with their surfactant-free counterparts. A distinct pattern of behaviour consistently emerges across the spectrum of $Bo$, clearly displaying the wave retardation phenomenon discussed in figure \ref{fig_press_vel_surf}. Simultaneously, a nonlinear trend in the $\tilde{\theta}$ vs. $\tilde{t}$ function is apparent in our surfactant cases (refer to the black dashed lines in each plot and the corresponding $R_{lin}^{2}$ values of a linear fit detailed in table \ref{tab:wave_vel_surf}). It is worth noting that our surfactant plots depict the last recoverable $\tilde{\theta}$ value before the singularity occurred in the simulations. This limitation arises due to the substantial pressures and velocities encountered, particularly near and at the moment of jet ejection, significantly complicating the accurate determination of the dominant wave's location. These challenges are further exacerbated in surfactant cases due to significantly larger local $Oh$, stemming from the high $\tilde{\Gamma}$ encountered at the wave focusing point ($T \approx 0$, refer to figure \ref{fig_press_vel_surf}(d,h) and figure \ref{fig_vel_wave_surf}(j)).

Figure \ref{fig_vel_wave_surf}(e)-(h) highlights a crucial feature of bursting bubbles with surface-active contaminants. The nonlinear nature of the functions $\tilde{\theta}$ vs. $\tilde{t}$ in surfactant cases does not allow data collapse by using $T$ and $\tilde{\theta}_{0}$, as was the case in clean interfaces for all $Bo$ and $Oh$ (see figure \ref{fig:vel_wave_clean}(c)). This non-collapsing behaviour between surfactant-free and surfactant-laden cases can be directly related to the findings of Ref. \cite{Pierre2022InfluenceBursting} in figure 2, where bubble collapsing times with surfactant could not be successfully rescaled by using $t_{c}$ to match those in clean interfaces. These observations, coupled with the insights of Ref. \cite{Constante-Amores2021DynamicsInterface} and the negligible influence of $Oh$ on wave velocity discussed in figure \ref{fig:vel_wave_clean}(a), collectively point towards Marangoni stresses—rather than a lower overall surface tension—as the primary cause for delays and nonlinearity in capillary wave motion. Further reinforcing this nonlinearity in surfactant cases, figure \ref{fig_vel_wave_surf}(i) and table \ref{tab:wave_vel_surf} demonstrate that the temporal evolution of $\tilde{\theta}$ can be more accurately described by a power-law of the form $\tilde{\theta} \sim \tilde{t}^{\alpha}$, where, generally, $\alpha$ increases with $Bo$, hinting at weakened surfactant effects with increased buoyancy effects.

To conclude this subsection, it is pertinent to comment on the moments nearing the collapse singularity in surfactant-laden cases. In Figure \ref{fig_vel_wave_surf}(k)-(l), we present our computed interfacial shapes for cases with insoluble surfactants across varying $Bo$, along with their scaled counterparts based on the self-similarity proposed in Ref. \cite{Lai2018BubbleProfiles} ($h(r,t) \sim |t_{0} - t|^{2/3}$, as discussed earlier in Figure \ref{fig:vel_wave_clean}(e)-(f) for clean cases). Interestingly, even in the presence of surfactant, we observe an almost perfect data collapse across time for each case below $Bo = 5$. This behaviour can be explained as follows: the times displayed in the plots align with the valid temporal ranges reported in Ref. \cite{Lai2018BubbleProfiles}, corresponding to $T \gtrapprox -0.1$. As can be seen in figure \ref{fig_vel_wave_surf}(j), the systems reach surfactant saturation at these comparative times, suggesting a local reduction (increase) in $Oh$ ($La$) and bringing them much closer to the collapsing $La$ range (1000 - 5000) reported in Ref. \cite{Lai2018BubbleProfiles} compared to the clean cases.

Although our power-law fits offer a good representation of the $\tilde{\theta}$ data, we emphasise that the complex dynamics in the collapse vicinity preclude us from tracking the dominant capillary wave all the way to $\tilde{\theta} = 0$, complicating the examination of the final stages. As previously mentioned, the curvature reversal at the instant of jet ejection disrupts the system's self-similar behaviour in clean cases as proposed in earlier works \cite{Lai2018BubbleProfiles,Duchemin2002JetSurface,Zeff2000SingularitySurface,Ghabache2014OnEjection}. In surfactant-laden cases, we also expect major disruptions to our proposed power-law $\tilde{\theta}$ behaviour as collapse is approached. In fact, we can infer a transition to a different regime, distinct from the power-law behaviour, at $T \gtrapprox -0.12$, given that that projections of the power-law fits to $\tilde{\theta} = 0$ do not match our recorded collapsing times in the simulations. For all cases, our recorded times fall behind those of the power-law projections, suggesting an acceleration of the dynamics. The authors recommend a thorough investigation of the moments very close to wave focusing and jet ejection ($\theta \approx 0$) in the presence of surfactants as an avenue for future research. Specifically, we recommend a focus on the ``Axe" shape adopted by the cavity as $\tilde{\theta} = 0$ is approached in the presence of surfactants, as seen previously in Refs. \cite{Pierre2022InfluenceBursting,Ji2023SecondarySurface}.

%
\begin{table}
\centering
\caption{Comparison of capillary wave velocity in clean interfaces and in the presence of insoluble surfactant at $\beta_{s} = 0.3$. $R^{2}_{lin}$ represents the coefficient of determination of the linear fits shown in figure \ref{fig_vel_wave_surf}(a-d) and $R^{2}_{pow}$ those of the power-law fits for the surfactant cases, shown in figure \ref{fig_vel_wave_surf}(i). These fits are of the form $\tilde{\theta} \sim \tilde{t}^{\alpha}$, where $\alpha$ is the best-fit exponent}
\label{tab:wave_vel_surf}
\begin{tabular}{c|ccc|ccccc}
\hline
\hline
 & \multicolumn{3}{c}{Clean} & \multicolumn{4}{c}{$|\tilde{\tau}_{m}| > 0$} \\
\,\,\,\, $Bo$ \,\,\,\, & \,\,\,\, $|\tilde{u}_{\theta}|$ \,\,\,\, & \,\,\,\, $|\tilde{u}_{\theta, lin}|$ \,\,\,\, & \,\,\,\, $R_{lin}^{2}$ \,\,\,\, & $|\tilde{u}_{\theta}|$ & \,\,\,\, $|\tilde{u}_{\theta, lin}|$ \,\,\,\, & \,\,\,\, $R_{lin}^{2}$ \,\,\,\, & \,\,\,\, $\alpha$ \,\,\,\, & \,\,\,\, $R_{pow}^{2}$ \,\,\,\, \\ \hline
$0.1$ & 5.228 & 5.145 & 0.998 & 3.202 & 2.967 & 0.961 & 0.456 & \,\,\,\, 0.997 \,\,\,\, \\
$0.5$ & 5.185 & 5.108 & 0.998 & 3.315 & 2.987 & 0.980 & 0.587 & \,\,\,\, 0.998 \,\,\,\, \\
$1$ & 4.844 & 4.951 & 0.999 & 3.125 & 3.003 & 0.975 & 0.517 & \,\,\,\, 0.998 \,\,\,\, \\
$5$ & 4.472 & 4.450 & 0.998 & 3.068 & 3.216 & 0.983 & 0.707 & \,\,\,\, 0.996 \,\,\,\, \\
\hline
\hline
\end{tabular}
\end{table}

\subsubsection{Effect of surfactant solubility and Marangoni stresses transition}\label{subsec_effect_Bi}

\begin{figure}
\includegraphics[width=1\linewidth]{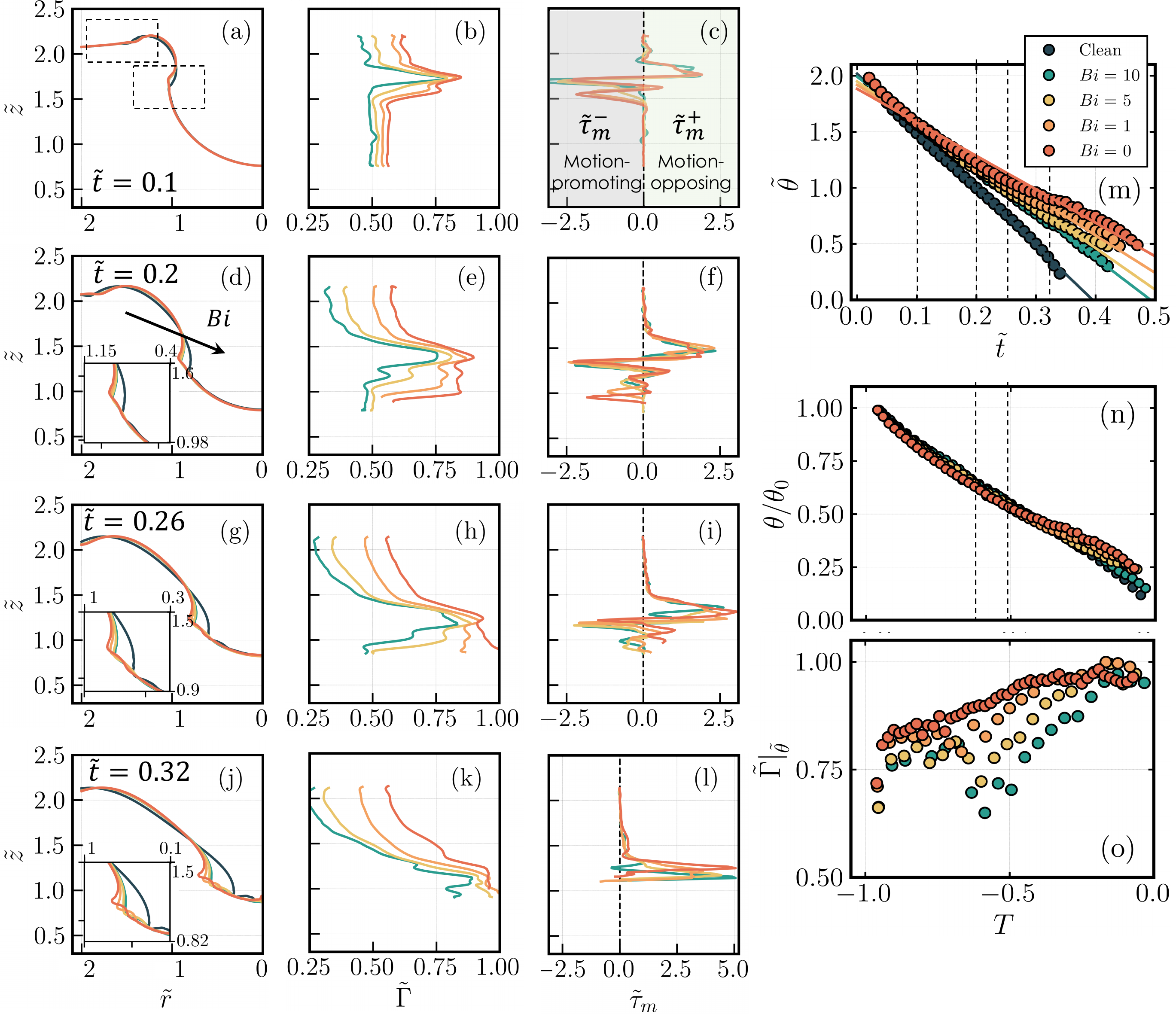}
  \caption{Effect of surfactant solubility on capillary wave dynamics: (a,d,g,j) interface shape; (b,e,h,k) surfactant concentration along the interface; (c,f,i,l) Marangoni stresses along the interface. In (a-l), the 
  abscissa 
  represent the values of $\tilde{r}$, $\tilde{\Gamma}$, and $\tilde{\tau}_{m}$, shared across each column, and the ordinate represent the vertical coordinate, $\tilde{z}$, shared across all plots; (m) temporal evolution of $\tilde{\theta}$ wherein the vertical dashed lines correspond to the times depicted in (a-l); (n) data scaled by $\theta_{0}$ and $\tilde{t}_{0}$; (o) interface surfactant concentration evaluated at the location of the dominant capillary wave. The markers represent our simulation data (taken every $\Delta \tilde{t} = 0.01$) and the lines correspond to a linear fit of the data for reference. For all cases in this figure, $Bo = 0.5$ and $Oh = 3.21 \times 10^{-3}$}
  \label{fig_vel_wave_Bi}
\end{figure}

Having dived into the general effects of surfactant addition, we now proceed to elucidate the effects of surfactant solubility and unveil the specific effects of Marangoni stresses on wave propagation. Figure \ref{fig_vel_wave_Bi}(a)-(l) presents a few snapshots of interface shape, $\tilde{\Gamma}$ profiles, and Marangoni stresses profiles in the $z$-direction, $\tilde{\tau}_{m}$, calculated via Eq. (\ref{eq:marangoni}) for various $Bi$. Rows 1-4 in this series of plots represent different dimensionless times and columns 1-3 display projections of $\tilde{r}$, $\tilde{\Gamma}$, and $\tilde{\tau}_{m}$ (shown in the $x-$axis) across the vertical axis, $\tilde{z}$ (shown in the $y-$axis). We reiterate here that under our formulation, an increase in $Bi$ signifies a higher likelihood of surfactant desorption into the bulk phase due to the smaller desorptive time-scale compared to that of convection. Consequently, the $Bi$ range is confined between a surfactant-free case ($Bi \to \infty$) and a fully insoluble surfactant ($Bi = 0$).

A clear monotonic trend regarding wave retardation and surfactant solubility is revealed, whereby lower $Bi$ (lower solubility) cases consistently exhibit more pronounced delays in capillary wave motion in both the train evolving down the cavity and the one departing from it. A comparison of interfacial shapes at early ($\tilde{t} = 0.1$) and later ($\tilde{t} = 0.32$) bursting stages suggests that these delays intensify in time and that a distinct separation between surfactant and clean cases emerges over time. This observation is consistent with those of Ref. \cite{Vega2024InfluenceBursting}, where even small surfactant concentrations ($C/CMC = 0.1$) were sufficient to actively alter the bursting dynamics. A measure of these delays is illustrated in figure \ref{fig_vel_wave_Bi}(m), where progressively larger deviations from the linear behaviour of the clean case arise as solubility is decreased. Similar to the insoluble cases at different $Bo$ presented in §\ref{subsec_surf_effect_Bo}, surfactant cases display non-collapsing dynamics against surfactant-free interfaces, as evidenced in figure \ref{fig_vel_wave_Bi}(n). Despite the higher rates of surfactant desorption in larger $Bi$ cases, reflected in the lower $\tilde{\Gamma}$ of the dominant wave at early and middle stages (figure \ref{fig_vel_wave_Bi}(o), $T < -0.25$), it is interesting to note that all cases reach surfactant saturation near collapse ($\tilde{\Gamma} \to 1$). This is attributed to the much smaller interfacial areas involved in the temporal vicinity of collapse (see for example figure \ref{fig_energy_clean}(b) at $T = 0$).

It is instructive to inspect the evolution of Marangoni stresses and surfactant concentration across the bubble's domain in the $z$-direction. The $\tilde{\Gamma}$ peaks at early stages ($\tilde{t} = 0.1 - 0.2$, figure \ref{fig_vel_wave_Bi}(b,e)), besides highlighting the location of the dominant wave, act as an important source of Marangoni stresses due to the generation of surface tension gradients (see Eq. (\ref{eq:marangoni})). In our formulation, positive-signed Marangoni stresses projected onto the $z$-direction ($\tilde{\tau}_{m} > 0$, denoted as $\tilde{\tau}_{m}^{+}$) are exerted in the positive $z$-direction, diverting flow away from the bottom of the bubble and wave focusing point. In contrast, negative Marangoni stresses ($\tilde{\tau}_{m} < 0$, denoted as $\tilde{\tau}_{m}^{-}$) tend to pull the interface towards the focusing point at $\tilde{r} = 0$. Armed with these definitions, it is clear that the high $\tilde{\Gamma}$ toroidal structure surrounding the dominant capillary wave engenders a complex interplay between positive and negative $\tilde{\tau}_{m}$, which retard and promote wave motion down the cavity, respectively. This is illustrated in figure \ref{fig_vel_wave_Bi}(c,f,i), in which two primary competing $\tilde{\tau}_{m}$ peaks are displayed. We henceforth refer to interfaces in this situation as subjected to ``bi-directional" Marangoni stresses. As the system advances towards surfactant saturation in the region surrounding the collapsing point ($\tilde{t} = 0.32$, figure \ref{fig_vel_wave_Bi}(j,k,l)), the motion-promoting negative Marangoni stresses nearly vanish, leaving the interface exposed to strong motion-delaying positive Marangoni stresses (note the $x$-axis scale difference in the $\tilde{z}$ vs. $\tilde{\tau}_{m}$ plots between $\tilde{t} = 0.32$ and the preceding times).

The mechanism outlined above elucidates what the authors define as a transition from bi-directional to uni-directional Marangoni stresses, a pivotal factor in the observed retardation of capillary waves on surfactant-laden bursting bubbles, as documented in this and prior studies. A visual depiction of this transition is provided in figure \ref{fig_press_vel_surf}(a)-(d). As shown in figure \ref{fig_vel_wave_Bi}(n), the magnitude of motion-delaying positive Marangoni stresses appears to overcome that of motion-promoting negative stresses within the time interval of $\tilde{t} = 0.20 - 0.26$. This critical period is demarcated by the corresponding vertical dashed lines in figure \ref{fig_vel_wave_Bi}(m), marking the juncture where the $\tilde{\theta}$ versus $\tilde{t}$ data for $Bi = 0$ begin to diverge markedly from the behaviour observed in surfactant-free cases. In the scaled space outlined by $\theta / \theta_{0}$ and $T$ in figure \ref{fig_vel_wave_Bi}(n), these times correspond to $T \approx [-0.60,-0.50]$ for $Bi = 0$, also coinciding with the onset of significant non-collapsing behaviour.

\subsubsection{Effect of surfactant elasticity and capillary wave velocity compilation}

\begin{figure}
\includegraphics[width=1\linewidth]{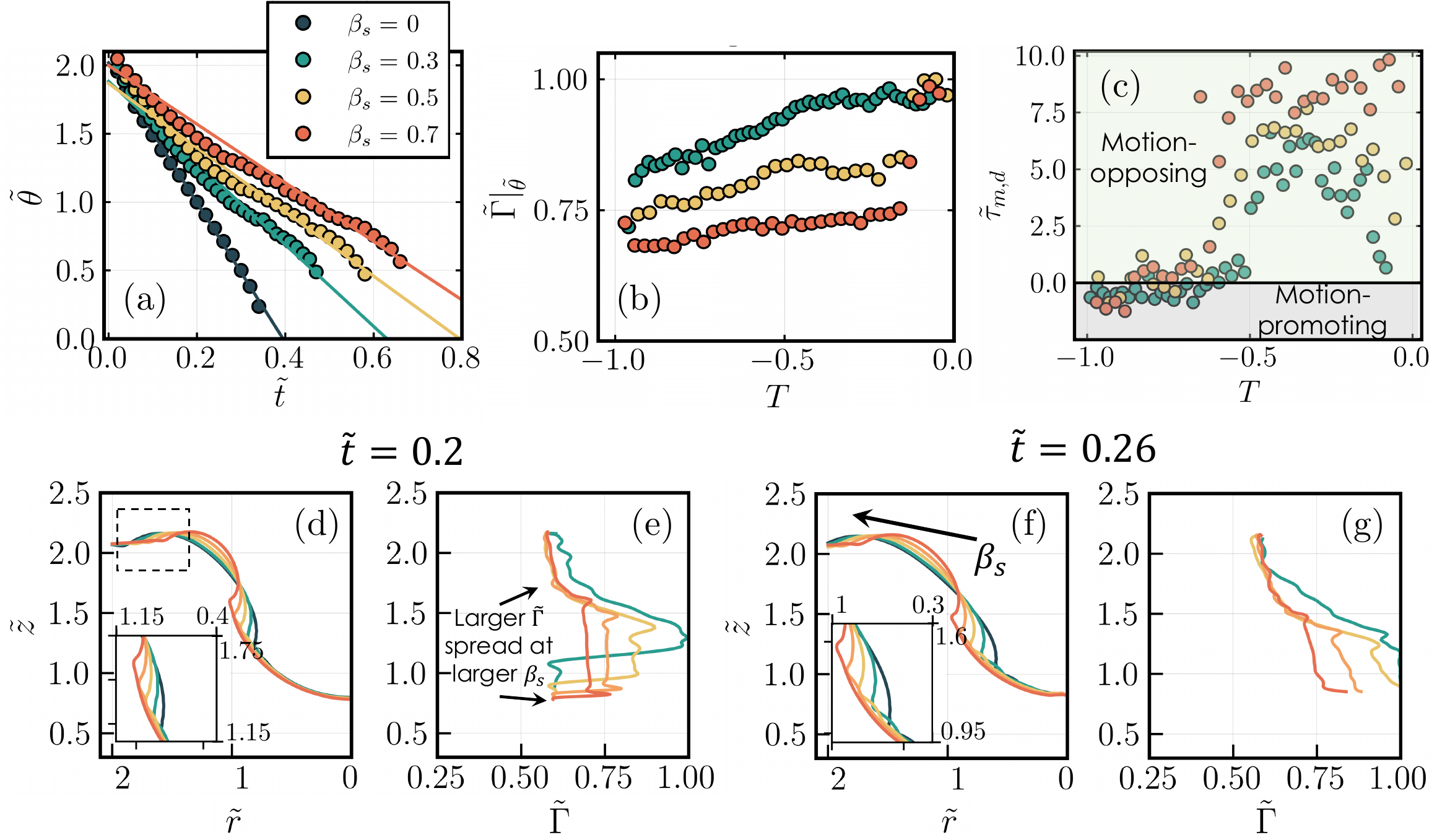}
  \caption{Effect of surfactant elasticity on capillary wave dynamics: (a) temporal evolution of $\tilde{\theta}$; (b) interface surfactant concentration evaluated at the location of the dominant capillary wave, scaled temporarily by the comparative time; (c) magnitude difference of positive (motion-opposing) and negative (motion-promoting) Marangoni stresses peaks across the interface; (d-g) sample interface shapes and surfactant concentrations along the interface for two different dimensionless times. In (d-g), the abscissa represent the values of $\tilde{r}$ and $\tilde{\Gamma}$ and the ordinate represent the vertical coordinate, $\tilde{z}$, shared across all plots. The markers represent our simulation data (taken every $\Delta \tilde{t} = 0.02$) and the lines correspond to a linear fit of the data for reference. For all cases in this figure, $Bo = 0.5$ and $Oh = 3.21 \times 10^{-3}$}
  \label{fig_vel_wave_beta}
\end{figure}

In this subsection, we illustrate the impact of surface elasticity, denoted by $\beta_{s}$, on wave propagation. Additionally, we present an analysis of the combined effects of $Bi$ and $\beta_{s}$ on average capillary wave velocity, $|\tilde{u}_{\theta}|$, across varying $Bo$ values. As explained in §\ref{subsec:dimless}, $\beta_{s}$ characterises the sensitivity of surface tension to surfactant concentration, as expressed in Eq. (\ref{eq:langmuir}). It also serves as a measure of the strength of Marangoni stresses, as seen in Eq. (\ref{eq:marangoni}). Figure \ref{fig_vel_wave_beta}(a) provides a comparison of the temporal evolution of $\tilde{\theta}$ for different values of $\beta_{s}$, with the surfactant-free case denoted as $\beta_{s} = 0$. Sample snapshots of the interface shape are also shown in figure \ref{fig_vel_wave_beta}(d,f). As anticipated, an increment in $\beta_{s}$ yields a positive, monotonic response in terms of wave motion retardation. Similar delays in capillary-related events with increased surfactant elasticity have been documented in previous studies concerning liquid thread retraction/breakup \cite{Constante-Amores2020DynamicsNumber,Craster2003PinchoffThreads}, ink-jet printing \cite{Antonopoulou2021EffectPrinting}, drop impact \cite{Constante-Amores2023ImpactFilms}, and bubble motion in capillary channels \cite{Pico2024DropChannels}.

It is interesting to note that the condition with the most pronounced wave delay (i.e., $\beta_{s} = 0.7$) does not align with the one exhibiting the highest $\tilde{\Gamma}$ at the dominant wave location at the same comparative time (refer to figure \ref{fig_vel_wave_beta}(b)). Similarly, it does not correspond to the largest $\tilde{\Gamma}$ in the toroidal region formed in the vicinity of the dominant wave, as seen in figure \ref{fig_vel_wave_beta}(e,g). This contrasts with the dynamics explored in our examination of $Bi$, where the case yielding the greatest delays ($Bi = 0$, insoluble surfactant) always corresponded to the highest $\tilde{\Gamma}$ at the dominant wave and its surrounding regions (refer to figure \ref{fig_vel_wave_Bi}(o) and (b,e,h,k)). This is easily explained by recalling the fact that an increase in $\beta_{s}$ is accompanied by an increase in Marangoni stresses strength. These stresses promptly redistribute $\tilde{\Gamma}$ across the interface, forming a larger surfactant `spread' near the dominant wave, leading to localised lower $\tilde{\Gamma}$ at larger $\beta_{s}$. These observations are in agreement with previous investigations on surfactant-laden bursting bubbles \cite{Constante-Amores2021DynamicsInterface}.

The preceding discussion offers a direct link to why larger values of $\beta_{s}$ precipitate lower wave propagation velocities. Illustrated in figure \ref{fig_vel_wave_beta}(c), we present our computations of the magnitude difference between the peaks of positive (motion-opposing) and negative (motion-promoting) Marangoni stresses in the $z$-direction at the beginning and end of the high $\tilde{\Gamma}$ toroidal structure, denoted as $\tilde{\tau}_{m,d} = |\tilde{\tau}^{+}_{m}| - |\tilde{\tau}^{-}_{m}|$. This parameter serves as a measure of the collective influence of Marangoni stresses within the system, with positive $\tilde{\tau}_{m,d}$ instigating overall wave motion retardation down the cavity. As depicted in figure \ref{fig_vel_wave_beta}(c), the time interval $T \approx [-0.75,-0.60]$ marks the juncture where motion-delaying stresses prevail over their motion-promoting counterparts as a whole. In this figure one can also see that the largest $\beta_{s}$ case consistently experiences the largest $\tilde{\tau}_{m,d}$ after $T \approx [-0.75,-0.60]$, confirming the presence of overall stronger motion-delaying Marangoni stresses and explaining the lower average wave velocities obtained for larger $\beta_{s}$ and their more pronounced departures from linearity.

\begin{figure}
\includegraphics[width=1\linewidth]{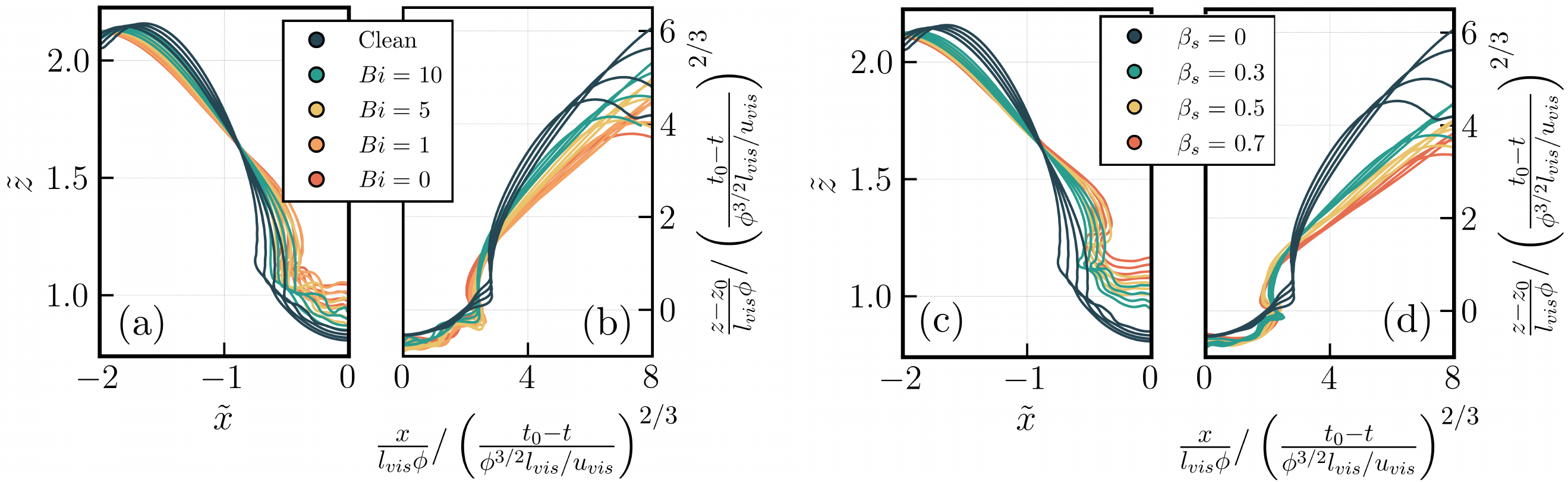}
\caption{Interface collapse under varying $Bi$ and $\beta_{s}$ conditions according to the scaling proposed in Ref. \cite{Lai2018BubbleProfiles} in the range $t_{0} - t \approx 6t_{pre} - 10t_{pre}$; $z_{0}$ represents the point in the vertical coordinate where jet ejection occurred; (a)-(b) effect of $Bi$ at constant $\beta_{s} = 0$; (c)-(d) effect of $\beta_{s}$ at constant $Bi = 0$. The clean case in (c)-(d) is denoted by $\beta_{s} = 0$. For these cases, $Oh = 3.21 \times 10^{-3}$ and $Bo = 0.5$. }
\label{fig_collapse_beta_Bi}
\end{figure}

In figure \ref{fig_collapse_beta_Bi}, we illustrate the effects of $Bi$ and $\beta_{s}$ at constant $Bo = 5$ and $Oh = 3.21 \times 10^{-3}$ on the universal interfacial shape profiles proposed in Ref. \cite{Lai2018BubbleProfiles}. As discussed in figures \ref{fig:vel_wave_clean}(e)-(f) and \ref{fig_vel_wave_surf}(k)-(l), a significant distinction is observed between the uncontaminated case (dark green) and the surfactant cases. This disparity is particularly noticeable in the $\beta_{s}$ plot in rescaled space, where the profiles for $\beta_{s} = 0.3 - 0.7$ collapse reasonably well over time and among themselves, contrasting sharply with the clean case which stands out as an outlier. As previously noted, the regions surrounding the bottom of the cavity in surfactant cases exhibit much larger $Oh$ values due to reduced surface tension, bringing the system closer to bursting conditions for which the scaling of Ref. \cite{Lai2018BubbleProfiles} was originally proposed. As demonstrated in figures \ref{fig_vel_wave_Bi}(a)-(l) and \ref{fig_vel_wave_beta}(d)-(g), Marangoni stresses, though significant in magnitude at these time points (see figure \ref{fig_vel_wave_beta}(c) for $T > -0.1$), primarily manifest upstream from the bubble's dominant wave, suggesting a Marangoni-free zone at the cavity's bottom and elucidating the better collapse observed in surfactant cases. We note here that the semi-collapsed interface profiles shown in figure \ref{fig_collapse_beta_Bi} correspond to times close to the bubble collapse singularity ($T > -0.1$), whose dynamics differ from the behaviour of $\tilde{\theta} vs. \tilde{t}$ discussed earlier.

\begin{figure}
\includegraphics[width=1\linewidth]{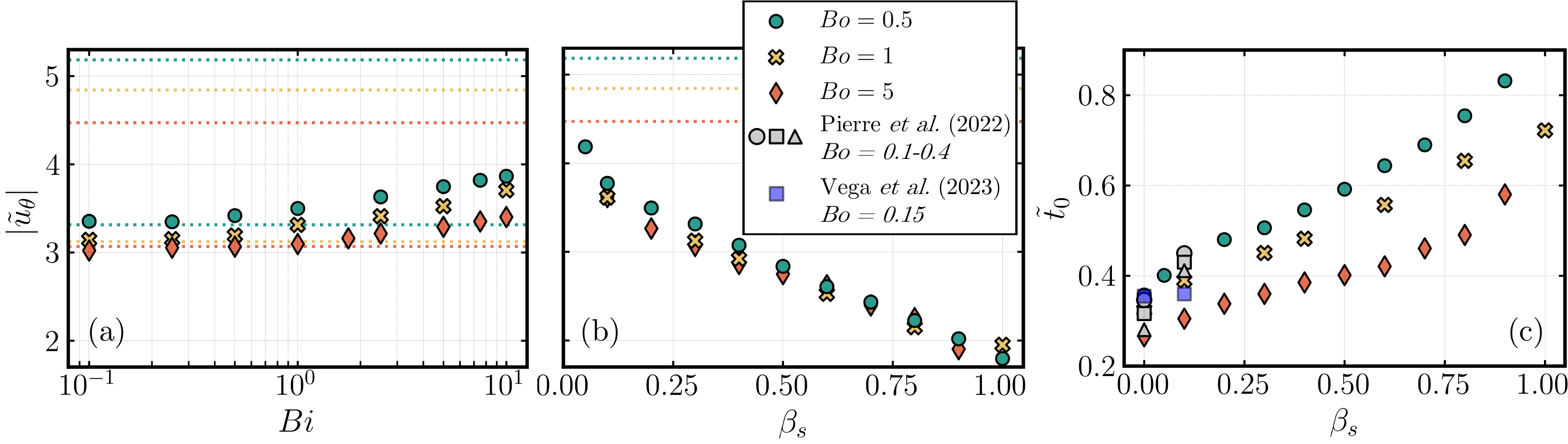}
\caption{Compilation of $Bi$ and $\beta_{s}$ effects on average dominant capillary wave velocity and collapse time under varying $Bo$ conditions: (a) effect of $Bi$ on $|\tilde{u}_{\theta}|$ at constant $\beta_{s} = 0.3$; (b) effect of $\beta_{s}$ on $|\tilde{u}_{\theta}|$ at constant $Bi = 0$. The dotted lines in (a) represent the average velocity of each $Bo$ for the clean case (top lines, $Bi \to \infty$) and insoluble surfactant (bottom lines, $Bi = 0$). The dotted lines in (b) represent the average velocity of each $Bo$ for the clean case ($\beta_{s} = 0$). For all cases in (a)-(b), $Oh = 3.21 \times 10^{-3}$; (c) effect of $\beta_{s}$ on collapse time and comparison against the experimental measurements of Refs. \cite{Pierre2022InfluenceBursting,Vega2024InfluenceBursting} ($Oh \sim O(10^{-3})$) at an equivalent equilibrium concentration $C/CMC \approx 0.25$}
\label{fig_effect_beta_Bi}
\end{figure}

We conclude our exploration of the pre-collapse dynamics by presenting an extended sweep of $Bi$ and $\beta_{s}$ in figure \ref{fig_effect_beta_Bi}(a)-(b), which illustrates the intertwined influence of these surfactant parameters on the average wave velocity, $|\tilde{u}_{\theta}|$, across varying $Bo$ conditions. It is relevant to reiterate that $|\tilde{u}_{\theta}|$ represents the \textit{average} velocity computed over time for each case, acknowledging the previously discussed nonlinear nature of $\tilde{\theta}$ vs. $\tilde{t}$ in surfactant-laden cases. Our extended parametric sweep confirms that the distinct monotonic effects of $Bi$ and $\beta_{s}$ elucidated previously persist up to our highest $Bo$ tested ($Bo = 5$). Specifically, we observe that $|\tilde{u}_{\theta}|$ exhibits a negative response (i.e., motion retardation) to a decrease in $Bi$ (less soluble surfactant) or an increase in $\beta{s}$ (`stronger' surfactant) for all $Bo$.

As previously discussed, the dependence of $|\tilde{u}_{\theta}|$ on $Bi$ and $\beta_{s}$ can be attributed to the significant increase in motion-delaying Marangoni stresses observed with decreasing $Bi$ or increasing $\beta_{s}$ (see, for example, figure \ref{fig_vel_wave_beta}(c)). We highlight, however, that these heightened stresses, which directly correlate with $\nabla_{s} \tilde{\Gamma}$ and $\beta_{s}$ (refer to Eq.(\ref{eq:marangoni})), are manifested differently for $Bi$ and $\beta_{s}$. In cases of decreased $Bi$, the higher $\nabla_{s} \tilde{\Gamma}$ at the start and end of the toroidal structure carrying the dominant wave induces the generation of larger Marangoni stresses compared to cases with larger $Bi$ (see Figure \ref{fig_vel_wave_Bi}(k)). Conversely, as $\beta_{s}$ increases, it is the greater magnitude of $\beta_{s}$ itself, rather than larger surface tension gradients at the toroidal structure, $\nabla_{s} \tilde{\Gamma}$, that leads to larger Marangoni stresses.

As illustrated by the identical ordinate range of the plots in figure \ref{fig_effect_beta_Bi}(a)-(b), $|\tilde{u}_{\theta}|$ appears to be consistently more sensitive to variations in $Bo$ across the tested range of $Bi$ compared to $\beta{s}$. This trend suggests that the influence of $\beta_{s}$ overwhelmingly dominates that of $Bo$. The non-negligible $Bo$ effects across our $Bi$ range align with the initial surfactant concentration effects on collapse time revealed experimentally by Ref.\cite{Vega2024InfluenceBursting}. In Figure \ref{fig_effect_beta_Bi}(c), a comparison of our dimensionless collapsing times against experimental data from Refs. \cite{Pierre2022InfluenceBursting,Vega2024InfluenceBursting} confirms that our results fall well within the ranges reported in previous investigations. Both studies referenced utilise sodium dodecyl sulphate (SDS), an insoluble surfactant characterised by $\beta_{s} \approx 0.1$, which is consistent with a Langmuir isotherm fit of the data presented in \cite{Vega2024InfluenceBursting}. Considering our initial interface surfactant concentration of $\tilde{\Gamma}_{0} = 0.55$, the approximately equivalent bulk concentration reported in the aforementioned references is $C/CMC = 0.25$, which corresponds to the data reported in the plot.

\subsection{Contaminated interfaces: jet drops and aerosol release}

\begin{figure}
\includegraphics[width=1.02\linewidth]{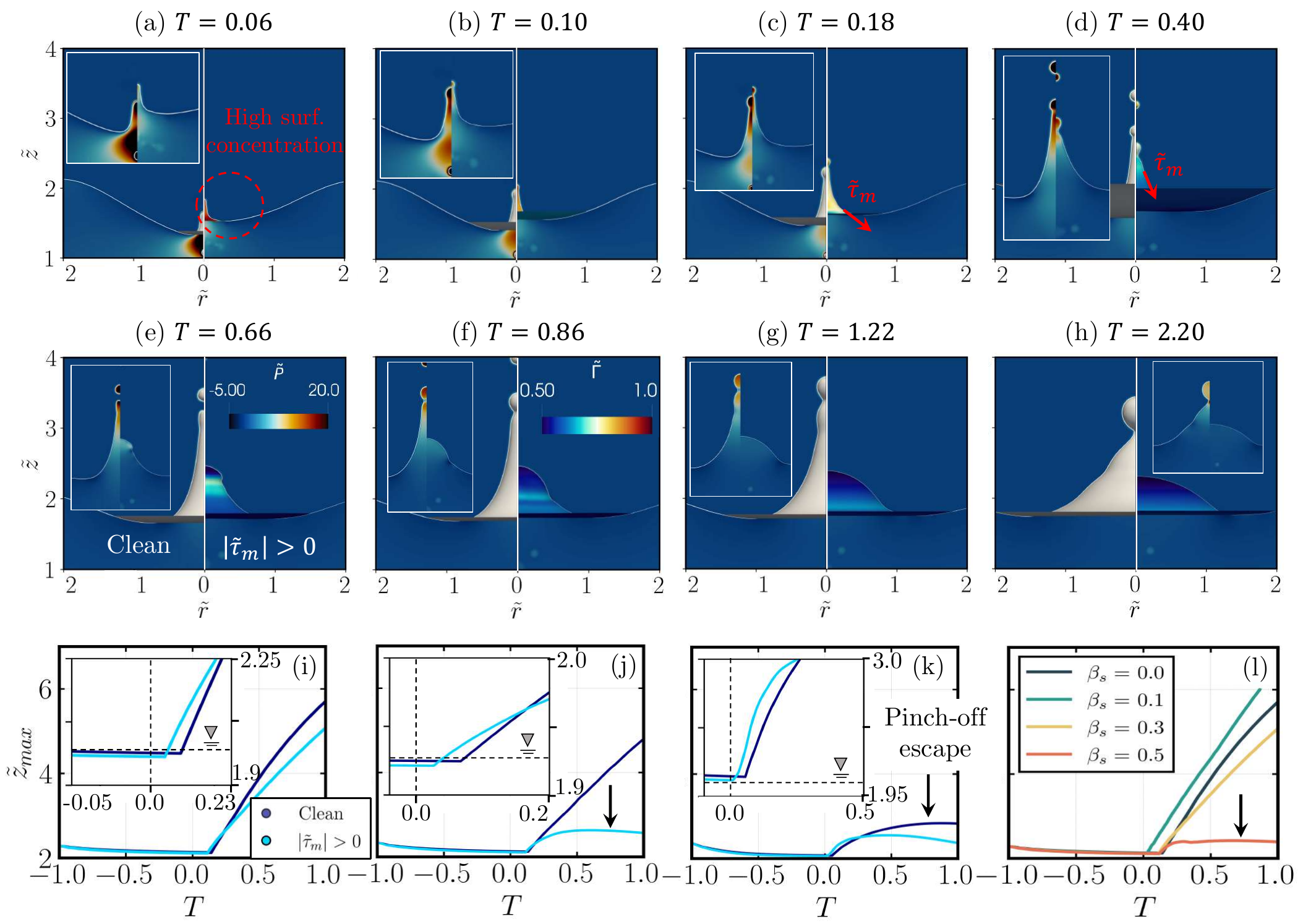}
  \caption{Effect of surfactant on Worthington jet dynamics: (a)-(h) comparison of temporal evolution of interface shape and pressure fields for surfactant-free interfaces on the left half and insoluble surfactant on the right half; the operating conditions for this case are identical to those in figure \ref{fig_press_vel_surf}. The insets focus on the emerging jet's pressure fields; (i)-(k) temporal evolution of maximal vertical location of the interface for $Bo = 0.5,1,5$, respectively; (l) effect of $\beta_{s}$ at $Bo = 0.5$.}
  \label{fig_jet_surf}
\end{figure}

We now focus briefly on the Worthington jet and the process of surfactant enrichment during collapse and jet breakup. As remarked in our introduction, recent works by Refs. \cite{Gordillo2023TheoryJets,Gordillo2019CapillaryJets} have proposed that the jet ejection phenomenon following cavity collapse is primarily driven by inertia, with viscosity playing a modulating role, rather than by an inertio-capillary mechanism. Figure \ref{fig_jet_surf}(a)-(h) presents a few snapshots of post-collapse interfacial shape and pressure contours for a surfactant-free (left half) and an insoluble surfactant case (right half). The parameters for these cases are identical to those in figure \ref{fig_press_vel_surf}. We highlight that these snapshots are taken at the same temporal distance from collapse and hence the visualisations represent the true differences between cases at the same stages of development. To complement these visualisations, we include plots of maximal interface location in the $z$-direction as a function of our comparative time for multiple $Bo$ and $\beta_{s}$ in figure \ref{fig_jet_surf}(i)-(l). As mentioned in our discussion of figure \ref{fig_energy_clean}(d), $\tilde{z}_{max}$ serves as a proxy for the location of the jet or first released drop.

Building upon the insights from §\ref{sec_surf_wave}, the observations from figure \ref{fig_vel_wave_Bi}(o) and figure \ref{fig_vel_wave_beta}(b), alongside previous work \cite{Constante-Amores2021DynamicsInterface}, demonstrate that surfactant concentration approaches saturation close to the moment of cavity collapse under our studied set of $Bi,\beta_{s}, Bo$. This critical observation lays the foundation for the emergence of Worthington jets in areas marked by high $\tilde{\Gamma}$, which, as a result, exhibit significantly larger local $Oh$ compared to cases devoid of surfactant contamination. Invoking the jet ejection velocity scalings proposed and validated by Refs. \cite{Gordillo2023TheoryJets,Gordillo2019CapillaryJets}, $\tilde{u}_{jet} \sim Oh^{1/2}$ for $Oh < 0.043$, or those of Ref. \cite{Ganan-Calvo2017RevisionSpeed}, $\tilde{u}_{jet}\sim Oh^{1/2}(Oh_{*} - Oh)^{-3/4}$ ($Oh_{*} = 0.043$), it is therefore expected that the addition of surfactant increases the dimensionless jet ejection velocity. This is evidenced qualitatively during the first few post-collapse stages in figure \ref{fig_jet_surf}(a)-(c) ($T < 0.18$), where the tip of the surfactant-covered jet consistently exceeds the vertical location of the surfactant-free jet. Examining the insets of figure \ref{fig_jet_surf}(i)-(k), with the horizontal dashed line representing the location of the liquid pool-atmosphere interface ($\tilde{z} = 2$) and the vertical line the bubble collapse time, it is clear that between the surfactant-free and surfactant-covered jet, it is the latter that first surpasses $\tilde{z} = 2$, suggesting a larger ejection velocity.

The authors of Ref. \cite{Pierre2022InfluenceBursting} have documented much larger jet velocities generated by surfactant-laden bursting bubbles prior to the first pinch-off event. We refer the reader to this publication for further insights into the influence of $La$ and surfactant presence on jet velocity during the first development stages. Interestingly, our results after these first ejection stages ($T > 0.18$, figure \ref{fig_jet_surf}(d)-(h)) uncover a complete trend reversal in terms of jet velocity and shape. While the surfactant-free jet continues to grow and undergo pinch-off events up until $T \approx 1.22$, the contaminated jet quickly reaches its maximum length at $T \approx 0.40$ and proceeds to drain out, leading to the thick, interfacially rigid structure shown at $T = 2.20$. These findings can be rationalised by incorporating the effects of Marangoni stresses into our analysis. As the jet grows in size, highlighted by the increase in interfacial area after $t_{col}$, surface tension gradients arise across the jet's domain, resisting the positive pressure gradient pushing the jet upwards and prompting interfacial stabilisation. This is schematically shown by the red arrows in the contour plots. An in-depth showcase of the Marangoni stresses profile along the Worthington jet can be found in Ref. \cite{Constante-Amores2021DynamicsInterface}.

The variations in $\tilde{z}_{max}$ resulting from changes in $\beta_{s}$ (figure \ref{fig_jet_surf}(l)) are a testament of the complex dynamics underpinning surfactant-laden bursting bubbles. In the case of $\beta_{s} = 0.1$, not only does the jet ejection velocity surpass that of the surfactant-free case, but its first ejected drop maintains this higher velocity throughout the entire temporal span. The above suggests that there exist situations in which Marangoni stresses originated from very weak surfactants cannot override the lowered capillary resistance to jet growth/breakup (larger local $Oh$) caused by the surfactant presence. Similar intricate surfactant behaviours have been reported in Ref. \cite{Pico2024DropChannels}.

In addition to jet velocity and size, surfactant contamination severely disrupts the capillary breakup dynamics of the Worthington jets. As is well-known in the literature across a wide array of fluid mechanics problems (see, for instance, Refs. \cite{Craster2003PinchoffThreads, Kamat2020Surfactant-drivenFilaments,Kamat2018RoleCascades, Liao2006DeformationMonolayer,Ambravaneswaran1999EffectsBridges,Constante-Amores2020DynamicsNumber,Kovalchuk2018EffectFormation,Kalli2023EffectApproach,Pico2024DropChannels,McGough2006RepeatedJet}), Marangoni stresses, rather than reduced local surface tension, play a primary role in seeking to reopen the capillary neck, thereby resisting end-pinching. In the context of bursting bubbles, this either dramatically reduces the number of ejected drops or completely inhibits the capillary singularity, as highlighted by the black arrows in figure \ref{fig_jet_surf}(j)-(l). In these cases, the combination of Marangoni-driven end-pinching resistance and the stabilising action of gravitational forces discussed in §\ref{subsec:results_clean} completely preclude the jet from pinching. We relate the aforementioned Marangoni-driven phenomenon to the `escape from pinch-off' mechanism in isolated liquid threads described by Refs. \cite{Kamat2020Surfactant-drivenFilaments,Constante-Amores2020DynamicsNumber}, wherein neck re-opening is attributed to the detachment and advection of a vortex ring towards the bulk of the forming bulbous tip.

To conclude our discussion, we delve into the process of contamination enrichment of aerosol drops. As demonstrated in previous investigations \cite{Dubitsky2023EnrichmentPosition}, any type of contamination in the liquid bulk can have highly problematic consequences due to the accumulation of these contaminants—oftentimes environmentally harmful—in the drops released to the atmosphere. Although the focus of this study comprised situations in which drop release is heavily suppressed, a few insights can be gathered from our simulations. Taking our base surfactant-free case, characterised by $Oh = 3.21 \times 10^{-3}$ and $Bo = 0.5$, a total of three ejected drops were obtained, with their radii relative to the bubble radius being $R_{d,i} / R_{0} = 7.7 \%$, $9.42 \%$, and $21.3 \%$, respectively ($i$ denotes the order of ejection). We note that the size of the first ejected drop agrees well with the reports by Ref. \cite{Blanchard1972ConcentrationBubbles} ($R_{d,1} / R_{0} \approx 6 \%$). In contrast, the sole drop produced by an insoluble surfactant at the same $Bo$ and a moderate $\beta_{s} = 0.3$ exhibited a size of $R_{d,1} / R_{0} = 4.9 \%$ and showed significant enrichment in terms of both interface and bulk surfactant concentration. Specifically, taking the surface average of $\Gamma$ and the volume average of $C$ within the drop, we obtain $\Gamma_{d,1} / \Gamma_{0} = 145.4 \%$ and $C_{d,1} / C_{\infty} = 774.4 \%$. This striking surfactant enrichment ($\approx 1.5 - 8$ times the initial values) is within the range of experimental enrichment factors of particulate and bacterial contaminants reported in Refs. \cite{Blanchard1972ConcentrationBubbles,Bezdek1972SURFACEBACTERIA,Ji2022Water-to-AirDrops}.


\section{Concluding remarks}\label{sec:conclusions}

We have dived into the classic phenomenon of bubbles bursting through a liquid-gas interface via numerical simulations, exploring two scenarios that, despite their practical significance, have only recently garnered attention: surface-active contamination and non-negligible gravitational forces compared to capillary forces, characterised by large Bond numbers, $Bo >> 10^{-3}$. Building upon the recent experimental \cite{Ji2023SecondarySurface, Vega2024InfluenceBursting, Pierre2022InfluenceBursting} and numerical \cite{Constante-Amores2021DynamicsInterface} work centred around surfactant effects on bursting bubble dynamics, our simulations have allowed us to provide detailed quantitative insights about the system. These include the temporal evolution of the dominant capillary wave angular location as it traverses down the cavity before collapse ($\tilde{\theta}$ vs. $\tilde{t}$), an exploration of the self-similarity of interfacial shape, Worthington jet growth, surfactant concentration and Marangoni stresses profiles along the interface, and the collective impact of surfactant parameters on these variables. This information is notably challenging to access experimentally. Herein, we have also proposed, for the first time, a phenomenological mechanism that explains the capillary wave retardation reported experimentally and numerically in surfactant-laden bursting bubbles.

Gravitational effects were found to be relevant in uncontaminated interfaces in terms of average velocity of the capillary wave, $|\tilde{u}_{\theta}|$, at large $Bo$. Prior investigations \cite{Blanco-Rodriguez2021OnBubbles} had reported insignificant variations in $|\tilde{u}_{\theta}|$ with $Bo$ for $Bo < 0.7$. Nonetheless, our study, which extended the examined Bond number range up to $Bo = 5$, reveals small but non-negligible decreases in $|\tilde{u}_{\theta}|$ with $Bo$ for our $Oh$ range considered, $1 \times 10^{-3} < Oh < 1 \times 10^{-2}$, suggesting a power-law fitting of the form $|\tilde{u}_{\theta}| \sim - Bo$. We have attributed the negative response of $|\tilde{u}_{\theta}|$ to $Bo$ to the much larger pressure gradients between the depths of the liquid pool and the areas surrounding the bubble at larger $Bo$, which compete with capillary wave motion and stabilise the system. In line with previous work for $Bo \to O(10^{-3})$ \cite{Gordillo2023TheoryJets,Gordillo2019CapillaryJets}, our extended $Bo$ range has confirmed that alterations in $Oh$ do not produce variations in $|\tilde{u}_{\theta}|$, even at $Bo = 5$. 

An important conclusion that can be drawn from this investigation corresponds to the substantially different pre-bubble collapse phenomena that emerge as surfactant is introduced to the system. Previous investigations, alongside our present research, have demonstrated that $\tilde{\theta}$ vs. $\tilde{t}$ behaves linearly in interfaces devoid of surfactant, implying constant $|\tilde{u}_{\theta}|$. Our simulations have revealed that Marangoni stresses, generated by non-homogeneous surface tension distributions, not only delay wave motion by about $30 - 70 \%$, but also completely disrupt the linearity of $\tilde{\theta}$ vs. $\tilde{t}$. For insoluble surfactants, dominant capillary wave motion was found to lean closer to $\tilde{\theta} \sim \tilde{t}^{\alpha}$, with $\alpha = 0.45 - 0.70$, depending on $Bo$. We note, however, that the dominant capillary wave undergoes a behavioural transition close to the instant of collapse, briefly exhibiting motion acceleration.  A comprehensive examination of the moments leading up to bubble collapse is left for future work.

Throughout this investigation we have introduced a mechanism that rationalises the delays in wave motion induced by surfactants, involving the transition from primarily `bi-directional' to primarily `uni-directional' Marangoni stresses. This mechanism can be summarised as follows. The large curvature that accompanies the surrounding areas of the dominant wave generates a toroidal region covered with surfactant (see figure \ref{fig_press_vel_surf}(a)) from which `downwards' ($\tilde{\tau}_{m} < 0$, motion-promoting) and `upwards' ($\tilde{\tau}_{m} > 0$, motion-opposing) Marangoni stresses emerge. As collapse is approached, surfactant saturation at the bottom of the bubble (see figure \ref{fig_press_vel_surf}(c)) eliminates $\tilde{\tau}_{m} < 0$, subjecting the interface to strong, uni-directional stresses heavily opposing wave motion across the cavity. This work has also presented a parametric sweep of two dimensionless groups relevant to various types of surfactants: surfactant solubility, given by $Bi$, and surfactant strength, given $\beta_{s}$. The former quantifies the rate at which surfactant desorbs into the liquid bulk phase while the latter is a proxy for the strength of Marangoni stresses. More pronounced reductions in capillary wave velocity were found for surfactants featuring weaker solubility (lower $Bi$) and larger $\beta_{s}$
This trend is explained by larger Marangoni stresses prompted by smaller $Bi$ and larger $\beta_{s}$. 

Finally, this study has complemented the numerical observations of Refs. \cite{Constante-Amores2021DynamicsInterface, Pierre2022InfluenceBursting} regarding surfactant effects on the Worthington jet. The unequivocally larger local $Oh$ close to the bubble collapse point brings about larger jet ejection velocities in surfactant cases, in line with the well-established scalings proposed in previous works for clean interfaces \cite{Ganan-Calvo2017RevisionSpeed,Gordillo2019CapillaryJets}. Nevertheless, jet expansion and growth force the emergence of strong Marangoni stresses that resist interfacial elongation and jet breakup via the Rayleigh–Plateau instability, ultimately leading to low-velocity jets and a significantly reduced number of ejected drops.

This investigation provides additional evidence to underscore the profound impact of surface-active contaminants, which are ubiquitous in nature, on the dynamics of bursting bubbles throughout the pre- and post-collapse stages. These effects extend beyond the mere retardation of bubble collapse, influencing the size distribution and degree of contamination of aerosol drops. As for future avenues of research, the authors suggest a numerical exploration of surface rheological effects with surfactants, a factor believed to significantly influence bursting bubbles as indicated by experiments of Ref. \cite{Ji2023SecondarySurface}. \\

\textbf{Declaration of Interests}. The authors report no conflict of interest.\\ 

\textbf{Acknowledgements and funding}. This work is supported by the Engineering and Physical Sciences Research Council, United Kingdom, through the EPSRC MEMPHIS (EP/K003976/1) and PREMIERE (EP/T000414/1) Programme Grants, and by computing time at HPC facilities provided by the Research Computing Service. O.K.M. acknowledges funding from PETRONAS and the Royal Academy of Engineering for a Research Chair in Multiphase Fluid Dynamics. We acknowledge support through HPC/AI computing time at the Institut du Developpement et des Ressources en Informatique Scientifique (IDRIS) of the Centre National de la Recherche Scientifique (CNRS), coordinated by GENCI (Grand Equipement National de Calcul Intensif) Grant 2023 A0142B06721. P.P. acknowledges the doctoral scholarship from the Colombian Ministry of Science, Technology and Innovation, MINCIENCIAS. The authors acknowledge with gratitude C.R. Constante-Amores for inspiring this work and for all fruitful discussions.\\


\appendix

\section{Code validation}\label{subsec:validation}

\begin{figure}
\includegraphics[width=1.0\linewidth]{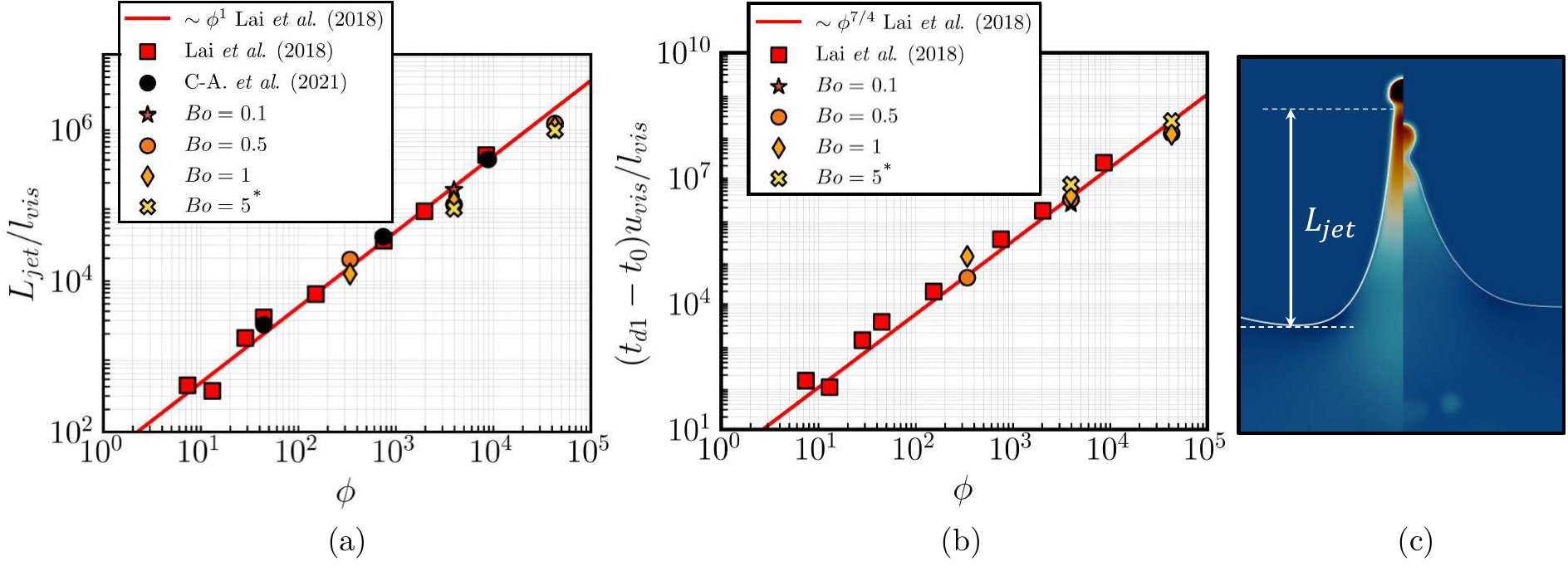}
  \caption{Comparison of our surfactant-free simulations for varying $Bo$ and $La$ against the scaling (red lines) and numerical simulations (red squares) of \citet{Lai2018BubbleProfiles} and the numerical data of \citet{Constante-Amores2021DynamicsInterface} at $Bo \sim \textit{O}(10^{-3})$ (black circles) using our same numerical approach: (a) dimensionless jet length before first drop detachment, $L_{jet}/l_{vis}$; (b) dimensionless temporal distance between cavity collapse and first detachment, $(t_{d1} - t_{0})u_{vis}/l_{vis}$; (c) schematic representation of how $L_{jet}/l_{vis}$ is measured. $^*$Measurements at the point of largest $L_{jet}$ attained for $Bo = 5$ given the complete escape from pinch-off.}
  \label{fig:val_lai}
\end{figure}

Due to the inherent complexity of bursting bubbles and their associated features, it is crucial to ensure that our numerical method and setup correctly recover the well-documented bursting behaviour at different stages in surfactant-free interfaces. In figure \ref{fig:val_lai}, we present our numerical results across different $Bo$ and $La$ conditions, comparing them with the scaling laws proposed by \citet{Lai2018BubbleProfiles} for two key parameters: jet length before first drop detachment, $L_{jet}$, and the temporal interval between first detachment, $t_{d1}$, and bubble collapse, $t_{0}$. These variables are made dimensionless by the viscous length- and time-scales ($l_{vis}$ and $t_{vis} \equiv l_{vis}/u_{vis}$, respectively); the x-axis variable, $\phi$, is defined in Eq. (\ref{eq:scales_viscosity}):

\begin{equation}\label{eq:scales_viscosity}
L_{vis} = \frac{\mu^{2}_{l}}{\rho_{l}\sigma_{s}}, ~~~
t_{vis} = \frac{\mu^{3}_{l}}{\rho_{l}\sigma^{2}_{s}} , ~~~
l_{vis} = \frac{\mu^{2}_{l}}{\rho_{l}\sigma_{s}} , ~~~
u_{vis} = \frac{\sigma_{s}}{\mu_{l}} , ~~~
\phi = La(La^{-1/2}_{*} - La^{-1/2}) = Oh^{-2}(Oh_{*} - Oh).        ~~~
\end{equation}

In the above, $La_{*} \approx 500$ denotes the critical value below which no drop detachment is observed as $Bo \to 0$. As depicted in figure \ref{fig:val_lai}, our simulation results exhibit favourable agreement with prior numerical and theoretical investigations. The minor deviations observed in terms of $L_{jet}$ with increasing $Bo$ indicate the presence of subtle but non-negligible gravitational effects beyond $Bo \approx 1$, impeding jet growth before the first detachment. In §\ref{subsec:results_clean}, we analyse the implications of these small effects, which become increasingly pronounced following initial detachment.

\begin{figure}
\includegraphics[width=1.0\linewidth]{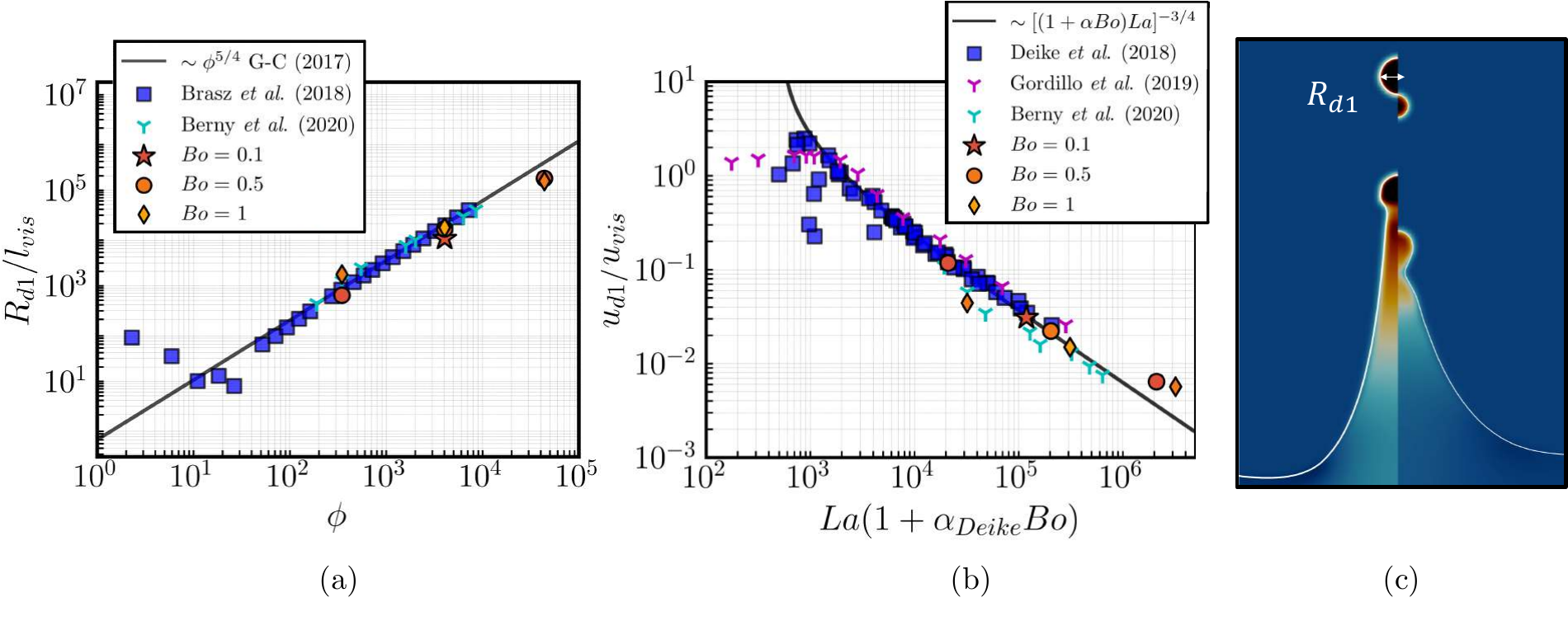}
  \caption{Comparison of the dynamics of the first released drop obtained with our surfactant-free simulations for varying $Bo$ against previous work: (a) scaling for dimensionless drop size, $R_{d1}/l_{vis}$, proposed by \citet{Ganan-Calvo2017RevisionSpeed} (black line); (b) $Bo$-dependent scaling for dimensionless drop velocity, $u_{d1}/u_{vis}$, proposed by \citet{Deike2018DynamicsBubbles} (black line). Here, $\alpha_{Deike} = 2.2$; (c) schematic representation of $R_{d1}$.}
  \label{fig_val_deike}
\end{figure}

We complement our validation by examining the size, $R_{d1}$, and velocity, $u_{d1}$, of the first ejected drop in figure \ref{fig_val_deike}. Drop sizes are compared against the scaling proposed by \citet{Ganan-Calvo2017RevisionSpeed}, while velocities are assessed against \citet{Deike2018DynamicsBubbles}'s extension of \citet{Ganan-Calvo2017RevisionSpeed}'s original law to include the effects of $Bo$. We also incorporate a few numerical and experimental results from previous investigations in the figures. Particularly, we include the experimental measurements of $R_{d1}$ by \citet{Brasz2018MinimumBubble} at $Bo = 0 - 1$ and the numerical results reported in \citet{Berny2020RoleBubbles} at $Bo = 1$. For $u_{d1}$, we display the results of \citet{Gordillo2019CapillaryJets} at $Bo = 5 \times 10^{-1}$, those of \citet{Deike2018DynamicsBubbles} at $Bo = 10^{-3} - 0.5$, and those of \citet{Berny2020RoleBubbles} at $Bo = 1$. As discussed in §\ref{sec:introduction}, $R_{d1}$ and $u_{d1}$ have been extensively investigated, resulting in a substantial body of numerical and experimental data validating the proposed scalings. Our figure includes a select set of datasets under conditions similar to ours for reference, but we encourage readers to consult Refs. \cite{Blanco-Rodriguez2021OnBubbles,Ghabache2016SizeBursting,Duchemin2002JetSurface,Seon2017EffervescenceEvaporation,Spiel1995OnSurfaces,Spiel1997MoreSurfaces,Krishnan2017OnSurface} for additional data points.

Ejected drop and Worthington jet velocities have been a subject of considerable debate, primarily due to measurement inconsistencies across different studies. In \citet{Deike2018DynamicsBubbles}, velocities are measured at the instant before jet breakup, when the jet has stabilised at a constant vertical velocity. This definition deviates from the approach of \citet{Gordillo2023TheoryJets} and collaborators, where the jet's velocity at its very inception—the maximum possible velocity in the system—is the variable considered and related to $La$. Others \cite{Ghabache2016SizeBursting}, perform their measurements at the moment the jet crosses the liquid pool-atmosphere interface. These inconsistencies explain why, for instance, \citet{Gordillo2019CapillaryJets}'s velocities in our plot are generally larger than \citet{Deike2018DynamicsBubbles}'s, even at similar $Bo$. By adhering to the latter reference's definition, we demonstrate how our simulations align well with the scalings observed in previous work. The point corresponding to $La = 10^{4}$ (shown as $La(1 + \alpha_{Deike}Bo) = 3.1 \times 10^{4}$) and $Bo = 1$ appears to be an outlier as our prediction for this points deviates significantly more than the rest, but remains within the same deviation order as the points of \citet{Berny2020RoleBubbles} at $Bo = 1$.


\section{Cap retraction and its effects of ejected jets}\label{subsec:cap}

\begin{figure}
\includegraphics[width=1.0\linewidth]{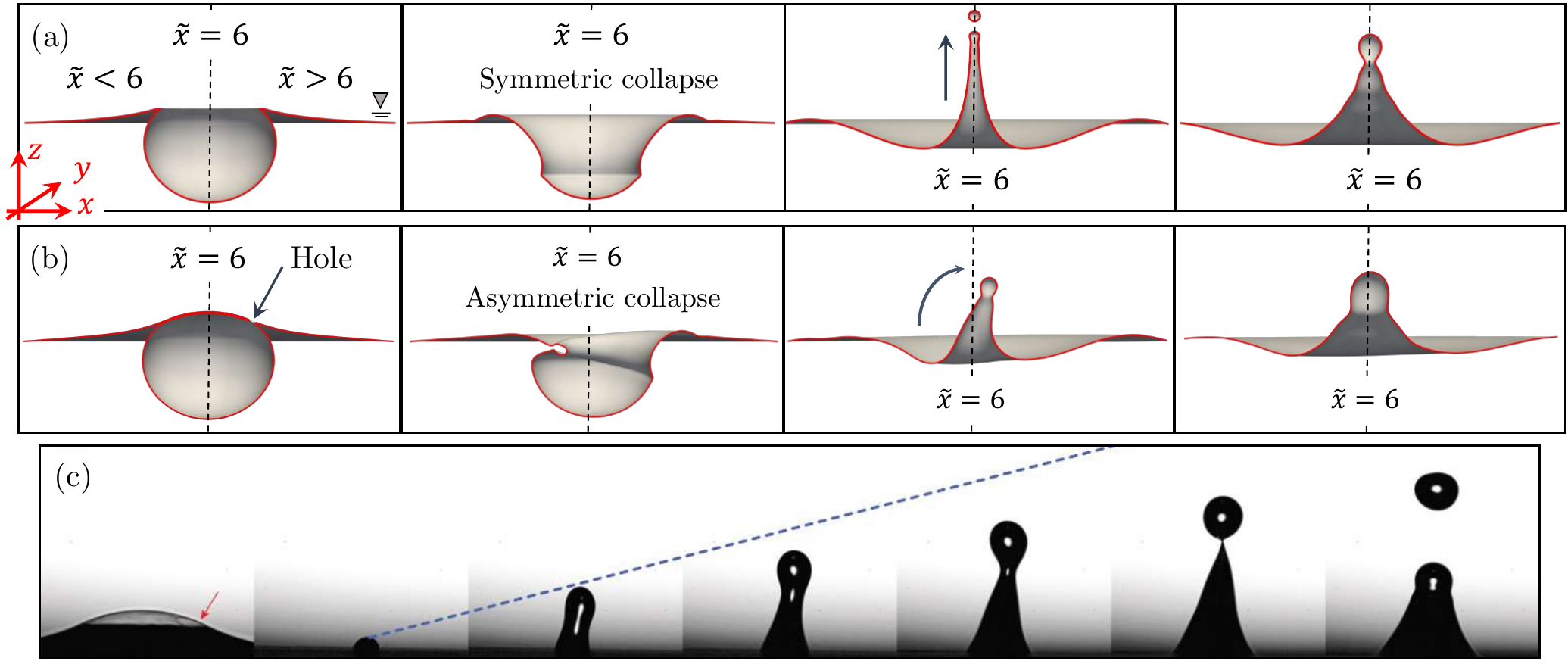}
  \caption{Influence of bubble cap and hole opening location on Worthington jet: (a) schematic of cap-less bursting, inducing a symmetric train of capillary waves along the bubble and generating a symmetric Worthington jet; (b) schematic of asymmetric bursting, generating an oblique jet. The simulation conditions of (a)-(b) correspond to $Bo = 0.5$ and $Oh = 1 \times 10^{-2}$ with cap thickness $\delta = 0.05R_{0}$; (c) experimental images of asymmetric bursting taken and adapted from \cite{ElisabethGhabache2015SurfaceEtires}}
  \label{fig_oblique_jet}
\end{figure}

As briefly mentioned at the end of §\ref{subsec:results_clean}, cap retraction dynamics are increasingly relevant as $Bo$ grows due to the comparable time-scales associated with cap retraction and bubble collapse. In all simulations, we opted for neglecting the effects of the cap given that the time associated with its retraction is typically at least 2-3 orders of magnitude lower than those of cavity collapse. Nevertheless, we highlight in this appendix that an asymmetric puncture in the cap can have enormous consequences in the pre- and post-collapse dynamics. As depicted in figure \ref{fig_oblique_jet}(a)-(b), a hole opening in the $\tilde{x} > 6$ region brings about the generation of an oblique jet due to the dominant capillary waves focusing at a point shifted away from $\tilde{x} = 6$. These phenomena have been observed experimentally in the work of Ref. \cite{ElisabethGhabache2015SurfaceEtires}, showcased in figure \ref{fig_oblique_jet}(c).

\bibliographystyle{plainnat}
\bibliography{references_burstingBubble.bib}

\end{document}